\documentclass[aps,reprint,groupedaddress,showpacs,floatfix]{revtex4-2}
\pdfoutput=1
\usepackage[utf8]{inputenc}
\usepackage[T1]{fontenc}
\usepackage{mathtools}
\usepackage{bm}
\usepackage[unicode]{hyperref}
\usepackage{siunitx}
\usepackage{upgreek}
\usepackage{xspace}
\usepackage{placeins}
\usepackage[caption=false]{subfig}
\usepackage[svgnames]{xcolor}
\usepackage[normalem]{ulem}

\DeclareMathOperator{\diag}{diag}
\DeclareMathOperator{\real}{Re}
\DeclareMathOperator{\imag}{Im}
\DeclareMathOperator{\Tr}{Tr}

\newcommand{\upe}{\ensuremath{\mathrm{e}}}
\newcommand{\upi}{\ensuremath{\mathrm{i}}}
\newcommand{\dif}{\ensuremath{\mathrm{d}}}
\newcommand{\kB}{\ensuremath{k_\mathrm{B}}}
\newcommand{\kBT}{\ensuremath{\kB T}}

\hypersetup{
    colorlinks=false,
    pdfauthor={W.P. Sterk},
    pdfsubject={elliptical magnon transport},
    pdftitle={Green's function formalism for nonlocal elliptical magnon transport},
    pdfkeywords={NEGF,
    magnon,
    ellipticity,
    anisotropy}
}

\newcommand{\ITP}{Institute for Theoretical Physics, Utrecht University,
Princetonplein~5, 3584~CC Utrecht, The~Netherlands\xspace}

\begin{document}
\author{W.P. Sterk}
\affiliation{\ITP}
\author{Andreas Rückriegel}
\affiliation{Institut für Theoretische Physik, Universität Frankfurt,
Max-von-Laue-Straße~1, 60438~Frankfurt, Germany\xspace}
\author{H. Y. Yuan}
\affiliation{\ITP}
\author{Babak Zare Rameshti}
\affiliation{Department of Physics, Iran University of Science and Technology,
Narmak, Tehran~16844, Iran\xspace}
\author{R.A. Duine}
\affiliation{\ITP}
\affiliation{Department of Applied Physics, Eindhoven University of
Technology, PO~Box~513, 5600~MB Eindhoven, The Netherlands\xspace}
\title{Green's function formalism for nonlocal elliptical magnon transport}
\date{\today}
\begin{abstract}
We develop a non-equilibrium Green's function formalism to study magnonic spin
transport through a strongly anisotropic ferromagnetic insulator contacted by
metallic leads. We model the ferromagnetic insulator as a finite-sized
one-dimensional spin chain, with metallic contacts at the first and last sites
that inject and detect spin in the form of magnons. In the presence of
anisotropy, these ferromagnetic magnons become elliptically polarized, and
spin conservation is broken. We show that this gives rise to a novel parasitic
spin conductance, which becomes dominant at high anisotropy. Moreover, the
spin state of the ferromagnet becomes squeezed in the high-anisotropy regime.
We show that the squeezing may be globally reduced by the application of a
local spin bias.
\end{abstract}

\maketitle

\section{Introduction}
The controllable transport of spin through magnetic materials has recently
attracted much attention, as it has the potential to augment or supplant
modern electronics with high-frequency and low-dissipation computational
elements \cite{zhang2014spintronics}. Various strategies have been envisioned
to achieve this goal, generally using either magnetic textures such as
skyrmions \cite{2018PhRvP...9f4018P,song2020skyrmion} or domain
walls \cite{2008Sci...320..190P,2019NatCo..10.4750V} as the carriers of
information, or using spin waves or magnons to transport spin angular momentum
directly. The latter forms a broad field of research known as
magnonics \cite{2010JPhD...43z4001K}. In recent years, significant milestones,
both experimental and theoretical, have been achieved in the field of
magnonics, with non-local transport of spin through ferromagnetic insulators
\cite{2015NatPh..11.1022C,2020PhRvP..13f1002F,2018PhRvL.120i7205W,guo2020nonlocal,2017PhRvB..95v4407P,2020AIPA...10a5031O}
now commonly realized and fairly well described using theoretical frameworks
that range from drift-diffusion models to non-equilibrium Green's function
formalism
\citep{2016PhRvB..94a4412C,2017PhRvB..96q4422Z,Nakata_2017,2019arXiv190302790U}.

At the core of these theoretical models is the Holstein-Primakoff (HP)
magnon \citep{1940PhRv...58.1098H}, a bosonic quasiparticle that forms a natural
approximation to low-energy excitations of the Heisenberg
(anti)ferromagnet \cite{1928ZPhy...49..619H}.
The simplest variants of the Heisenberg ferromagnet do not include any form of
anisotropy, or have at most a `natural' quantization axis, generally taken to
be the $z$ axis, set by an external magnetic field. This results in a
circularly polarized magnon, which appears to offer a sufficient approximation
to adequately describe the broad behavior of magnon transport
\cite{2012PhRvL.108x6601B}, for example in materials such as yttrium iron
garnet \cite{2020ApPhL.117i0501K}.

In this work, however, we explicitly consider the effects of potentially large
anisotropies, which break spin conservation and generate
elliptically polarized magnons. The breaking of spin conservation is known to
give rise to phenomena such as magnon tunneling between weakly coupled
ferromagnetic insulators \cite{2020PhRvB.101i4402Z}, which is prohibited when
spin is conserved, and super-Poissonian shot noise \cite{2020ApPhL.117i0501K}.
Such phenomena are expected to arise whenever the ferromagnet under
consideration has sufficiently strong anisotropy, e.g. in iron thin films
\cite{2020ApPhL.117i0501K} or exotic quantum magnets
\cite{2018PhRvL.120q7202U}.

We develop a non-equilibrium Green's function (NEGF) formalism, also known as
Keldysh formalism \cite{keldysh1965diagram,rammer2007quantum}, to study the
anomalous or off-diagonal correlations that are generated by the anisotropy
terms, and as a proof of concept, apply it to determine whether magnon
ellipticity gives rise to observable effects in local- and nonlocal transport
experiments.

We find that, given sufficiently strong anisotropy, at least two potentially
observable effects are produced: a novel parasitic spin resistance, and
phase-space squeezing of magnons. The parasitic spin resistance may provide
experimental insight into the anisotropy of the ferromagnet, provided a way
can be found to measure it directly. Squeezed magnons are predicted to yield
reduced shot noise in ferromagnet/conductor hybrids
\cite{2016PhRvL.116n6601K}, analogous to the application of squeezed light to
reduce quantum noise in optical lasers
\cite{1983Natur.306..141W,aggarwal2020room}. This effect may also
hypothetically find an application in the recently proposed magnon laser
\cite{2019PhRvL.122c7203D}.

The outline of this work is as follows: in Section~\ref{sec:methods}, we
recast the continuum field theory briefly outlined by
\citet{2020AnPhy.41268010R} into a discrete, $N$-spin form using a bottom-up
approach (similar work has been done in contexts such as the Bose-Hubbard
model \cite{2011PhRvA..84a3613G}), and in Section~\ref{sec:results}, show the
results we obtain from a numerical implementation of the framework. In
Section~\ref{sec:conc} we provide some concluding remarks and outline some
potential further applications of the formalism developed in this work.

\section{Methods}
\label{sec:methods}
In this section, we give a description of our model system and the
implementation of the NEGF we use to investigate its dynamics.

\subsection{System and Hamiltonian}
\label{subsec:sys}
\begin{figure}
    \centering
    \includegraphics[width=\linewidth]{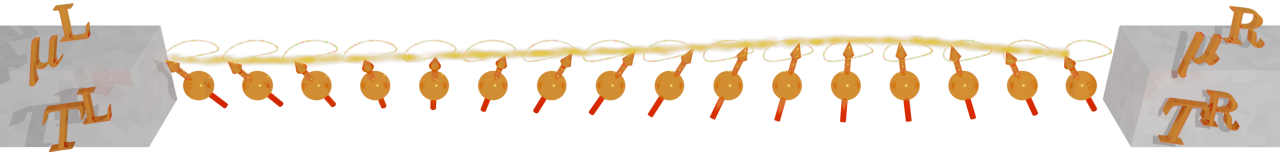}
    \caption{Cartoon representation of the system under consideration. A
    one-dimensional chain of spins is terminated at either end by heavy-metal
    leads, the left (right) lead having an electronic spin accumulation
    $\mu^\mathrm{L(R)}$ parallel to the magnetization axis, and temperature
    $T^\mathrm{L(R)}$. Spins precess elliptically due to the presence of high
    anisotropy, and can transport angular momentum in the form of an
    elliptical magnon or spin wave (yellow swirl).}
    \label{fig:spinwave1}
\end{figure}

We aim to consider systems typically used in long-distance transport
experiments akin to \citet{2015NatPh..11.1022C}: a ferromagnetic insulator
with two heavy-metal leads, one of which serves to inject magnons, and one
acting as a magnon detector. The system is biased by a constant electronic
spin accumulation in the leads, aligned parallel to the magnetization so that
there is no torque acting on the magnetization.  We assume spin transport in
the ferromagnetic insulator is quasi-one-dimensional, i.e., that magnons
travel in a straight line from emitter to detector, and the bulk of the
ferromagnet effectively consists of macroscopically many non-interacting
parallel copies of the spin chain making up the transport channel for a single
magnon. This allows us to treat the ferromagnetic bulk as a one-dimensional
(1D) spin chain. A cartoon representation of this system is shown in
Fig.~\ref{fig:spinwave1}.  Extension to a two- or three-dimensional cubic bulk
is mathematically simple (and tractable in the continuum limit), but
computationally challenging for finite-sized systems due to the vast increase
in lattice sites that must be taken into account.

We thus model our system using the 1D, $N$-particle Heisenberg
\cite{1928ZPhy...49..619H} ferromagnetic insulator in the presence of
quadratic anisotropy terms. It is described by the Hamiltonian\begin{align}
    H&=H_\text{H}+H_\text{ani}, \label{eq:hamfull}
    \intertext{where}
    H_\text{H}&=-\frac{\tilde{J}}{2}\sum_{i=1}^{N-1}\bm{\hat{S}}_i\cdot
            \bm{\hat{S}}_{i+1}-h_\mathrm{mag}\sum_{i=1}^N\hat{S}_i^z
    \intertext{is the ordinary 1D Heisenberg
    Hamiltonian \cite{1928ZPhy...49..619H}, and}
    H_\text{ani}&=\sum_{\nu\in\{x,y,z\}}
            \sum_{i=1}^N K_\nu(\hat{S}^\nu_i)^2
\end{align}is the anisotropy Hamiltonian. Here $\tilde{J}>0$ is the exchange
constant, the $K_\nu$ are the anisotropy energies in the three Cartesian
directions, $\hat{\bm{S}}_i=(\hat{S}^x_i, \hat{S}^y_i,
\hat{S}^z_i)^\mathrm{T}$ is the spin operator at site $i$, and
$h_\mathrm{mag}$ is an externally applied magnetic field.

As we are only interested in the behavior of the ferromagnet, we have omitted
Hamiltonian terms originating from coupling to the leads, and instead opt to
directly write down the relevant self-energy terms when we develop our Green's
function formalism later on.

The second-order, spin-$S$ Holstein-Primakoff transformation
\citep{1940PhRv...58.1098H}\begin{subequations}\label{eq:hptrans}\begin{align}
    \hat{S}^x_i&=\sqrt{\frac{S}{2}}\left(b_i+b^\dagger_i\right),\\
    \hat{S}^y_i&=-\upi\sqrt{\frac{S}{2}}\left(b_i-b^\dagger_i\right), \\
    \hat{S}^z_i&=S-b^\dagger_ib_i
\end{align}\end{subequations}
is used to express the Hamiltonian \ref{eq:hamfull} in terms of
magnon creation (annihilation) operators $b^\dagger_i$ ($b_i$) acting
at site $i$, that obey the bosonic commutation relations $[b_i,
b_j]=[b^\dagger_i,b^\dagger_j]=0$ and $[b_i, b^\dagger_j]=\delta_{ij}$.
We additionally define the vector operator\begin{align}
    \label{eq:phiops}
    \phi_i&\equiv\begin{pmatrix} b_i \\ b^\dagger_i\end{pmatrix}
\end{align}and its conjugate transpose $\phi^\dagger_i$.

Note that the Holstein-Primakoff transformation is an expansion around the
ground state in which all spins are aligned in the $z$-direction. In the
absence of an external field, this puts constraints on the relative signs and
strengths of the anisotropy terms $K_\nu$, however a sufficiently strong field
$h_\mathrm{mag}>0$ may always be used to guarantee alignment to the
$z$-axis.

The Hamiltion of Eq.~(\ref{eq:hamfull}) may be simplified somewhat if one
defines the constants $\Delta\equiv S(K_x+K_y-2K_z+\tilde{J})+h_\mathrm{mag}$,
$J\equiv\frac{\tilde{J}S}{2}$ and $K\equiv\frac{S}{4}(K_x-K_y)$. Then,
dropping unimportant constant energy shifts, along with the additional
boundary terms $-J(b^\dagger_1b_1+b^\dagger_Nb_N)$ that originate from the
fact that we consider a finite-sized system (which we expect to be negligible
for sufficiently large systems), the Hamiltonian (\ref{eq:hamfull}) may be
rewritten as\begin{align}
    H&=\frac{1}{2}\sum_{ij}\phi^\dagger_ih_{ij}\phi_j, \label{eq:Ham}
    \intertext{with the $2N\times2N$ matrix}
    \label{eq:hammat}
    h_{ij}&=\begin{pmatrix}h^\mathrm{i}_{ij}&K\delta_{ij}\\
        K\delta_{ij}&h^\mathrm{i}_{ij}\end{pmatrix}.
    \intertext{Here $\delta_{ij}$ is the $N\times N$ identity matrix, and}
    h^\mathrm{i}_{ij}&=\Delta\delta_{ij}
            -J\left[\delta_{i,j+1}+\delta_{i+1,j}\right]
\end{align}
is the isotropic Hamiltonian submatrix. We thus see that $\Delta$ is an
on-site potential for the magnons. The rescaled exchange energy $J$ is a
hopping parameter, governing the probability for a magnon to hop from one site
to the next.

The off-diagonal submatrices $K\delta_{ij}$ govern the ellipticity of the
magnons, and we shall henceforth use the term ``the anisotropy''
interchangeably with ``the scalar constant $K$'' (alternatively, and
equivalently, $K$ could be called the ``squeezing factor'' or ``spin
nonconservation factor''). Note, however, that $K$ is proportional to the
\emph{difference} in anisotropy energies in the $x$ and $y$ directions, i.e.
the principal directions perpendicular to the spin quantization axis.

The presence of nonzero $K$ breaks conservation of spin by introducing terms
of the form $K\big[b_ib_i+b_i^\dagger b_i^\dagger\big]$. The Hamiltonian of
Eq.~(\ref{eq:hammat}) therefore cannot be unitarily diagonalized (in a
physically meaningful way), and its eigenstates do not have a well-defined
spin. Rather, Eq.~(\ref{eq:hammat}) describes a Hamiltonian of the
\emph{Bogoliubov form}, which may be diagonalized using a para-unitary
transformation \citep{1978PhyA...93..327C}, i.e. a transformation matrix
$\mathcal{T}_{ij}$ obeying\begin{align}
    \sum_{i}(\sigma_3)_{ij}\mathcal{T}^\dagger_{jk}
        &=\sum_{i}\mathcal{T}^{-1}_{ij}(\sigma_3)_{jk},
    \intertext{where}
    (\sigma_3)_{ij}&\equiv\begin{pmatrix}
            \delta_{ij}&0\\0&-\delta_{ij}\end{pmatrix}
    \intertext{is the $2N\times2N$ analog of the third Pauli matrix (referred
    to as the para-identity matrix by \citet{1978PhyA...93..327C}). The
    Hamiltonian (\ref{eq:hammat}) allows us to choose $\mathcal{T}_{ij}$
    to be real, such that it takes the simple block structure}
    \mathcal{T}_{ij}&=\begin{pmatrix}
                \mathcal{T}^{(1)}_{ij} & \mathcal{T}^{(2)}_{ij}\\
                \mathcal{T}^{(2)}_{ij} & \mathcal{T}^{(1)}_{ij}\end{pmatrix},
\end{align}
where the individual $N\times N$ blocks $\mathcal{T}^{(1)}_{ij}$ and
$\mathcal{T}^{(2)}_{ij}$ are \emph{not} symmetric.

The para-unitary diagonalization of $h_{ij}$ is performed analytically for
arbitrary $N\ge2$ by leveraging the recurrent structure of the characteristic
equation $\det\{h_{ij}-(\sigma_3)_{ij}\varepsilon\}=0$, whereby the $N$-level
equation can be expressed terms of the $(N-1)$- and $(N-2)$-level equations.
The characteristic polynomial of the recurrence relation contains only terms
of degree $N+1$, and is therefore easily solved analytically. The
quasiparticles are elliptical magnons with the dispersion
relation\begin{align}
    \label{eq:dispersion}
    \varepsilon_n&=\sqrt{\left[\Delta-2J\cos\left(\frac{n\uppi}{N+1}\right)
            \right]^2-K^2}, &1&\le n\le N.
\end{align}
Here, the natural number $n$ is the quantum number, and $\varepsilon_n$
monotonically increases with $n$. The corresponding eigenstates are plane
waves \footnote{Specifically, the components of the paravector with quantum
number $n$ corresponding to site $i$ are simply $\sin\left(\frac{i n
\uppi}{N+1}\right)$, up to paranormalization.}, with the quantum number $n$
corresponding to the wavenumber $k=\frac{n\uppi}{L}=\frac{n\uppi}{Na}$ for a
spin chain of physical length $L=Na$, with $a$ the lattice constant.

\subsection{Non-equilibrium Green's function formalism}
\label{subsec:framework}
As stated in the previous subsection, diagonalization of our anisotropic
ferromagnetic insulator Hamiltonian may be done analytically and results in
free elliptical magnon modes. We now seek to investigate the
finite-temperature steady-state behavior of such a system in the presence of
two effects: (1) coupling to one or more metallic leads and (2) bulk
dissipation of elliptical magnons in the form of Gilbert-like damping.

To this end, we develop a non-equilibrium Green's function
framework \cite{datta2000nanoscale}, also known as Keldysh
formalism \cite{keldysh1965diagram,rammer2007quantum}. In what follows, we set
$\hbar=1$. The spectral properties of the magnons are encoded in the
single-particle retarded Green's function\begin{align}
    g_{ij}(t, t')&=-\upi\theta(t-t')\left<
            \left[\phi_i(t), \phi^\dagger_j(t')\right]\right>
    \intertext{and advanced Green's function}
    g^\dagger_{ij}(t, t')&=\upi\theta(t'-t)\left<
            \left[\phi_i(t), \phi^\dagger_j(t')\right]\right>,
    \intertext{where $\theta(t-t')$ is the Heaviside step function and
    $[\bullet,\bullet]$ is the commutator. The Keldysh Green's function}
    g^\mathrm{K}_{ij}(t, t')&=-\upi\left<
            \left\{\phi_i(t),\phi^\dagger_j(t')\right\}\right>
    \intertext{encodes information about the occupation of the single-particle
    states. Here, $\{\bullet,\bullet\}$ is the anticommutator. Using these
    Green's functions, one may construct the lesser Green's function}
    g^<_{ij}(t, t')&=-\upi\left<\phi^\dagger_i(t)\phi_j(t')\right>,
    \intertext{which, at equal times $t=t'$, contains the off-diagonal
    correlations (for $i\ne j$) and quasiparticle number density (for $i=j$),
    up to a prefactor of $-\upi$. Together with the greater Green's function}
    g^>_{ij}(t, t')&=-\upi\left<\phi_i(t)\phi^\dagger_j(t')\right>,
    \intertext{one obtains the relations \cite{rammer2007quantum}}
    g_{ij}(t, t')&=\theta(t-t')\left[
            g^>_{ij}(t, t')-g^<_{ij}(t, t')\right], \\
    g^\dagger_{ij}(t, t')&=-\theta(t'-t)\left[
            g^>_{ij}(t, t')-g^<_{ij}(t, t')\right], \\
    g^\mathrm{K}_{ij}(t, t')&=g^>_{ij}(t, t')+g^<_{ij}(t, t'),
\end{align}
and\begin{align}
    g_{ij}(t, t')-g^\dagger_{ij}(t, t')&=g^>_{ij}(t, t')-g^<_{ij}(t, t').
\end{align}

For simplicity, we shall henceforth drop the subscripts $i, j, \dotsc$ on all
matrices, as well as the explicit summations in matrix products seen in
Section~\ref{subsec:sys}, and work in the space of $2\times2$ matrices, of which
the four components are themselves $N\times N$ matrices. The presence of the
$N\times N$ or $2N\times2N$ identity matrix is implied when doing so does not
lead to ambiguity.

For the remainder of this work, we shall only consider a system in the steady
state, i.e. $g(t, t')=g(t-t')$ (and similar for the other Green's functions),
and work with Fourier-transformed Green's functions. In particular, the
retarded Green's function $g(\omega)$ satisfies the Dyson
equation\begin{align}
    \label{eq:dyson}
    g(\omega)&=[\omega\sigma_3-h-\Sigma(\omega)]^{-1},
\end{align}
where $\Sigma(\omega)$ is the retarded self-energy, and is easily obtained by
numerical matrix inversion.

Here, we opt to stay in the HP basis (i.e., the basis of the \emph{circular}
magnons defined by the operators $\phi$ and $\phi^\dagger$) instead of
transforming to the elliptical basis, and thus $h$ in Eq.~(\ref{eq:dyson}) is
simply given by Eq.~(\ref{eq:hammat}). Lead coupling and Gilbert-like damping
are to be incorporated into the (retarded) self-energy $\Sigma$.

The reason we choose to compute observables in the circular basis is twofold:
(1) it provides a simple form for the lead self-energies, which will be
explained shortly, and (2) experimental measurement of observables is
generally done electrically (through the spin Hall effect and its inverse)
\cite{2015NatPh..11.1022C,2016ApPhL.109b2405G}, so that electron spin is the
natural measurement basis (see below).

In line with \citet{2017PhRvB..96q4422Z}, we take the self-energy component
arising from lead $X$ to have the form\begin{align}
    \Sigma^X(\omega)&=-\upi\eta^X(\omega-\mu^X\sigma_3)
            \delta_{i,i^X}\delta_{j,i^X}, \label{eq:lead-se}
\end{align}where $\delta_{i,i^X}\delta_{j,i^X}$ indicates that the self-energy
is zero everywhere except for its diagonal components corresponding to site
$i^X$\footnote{Thus the full $2N\times2N$ matrix has nonzero components at
indices $(i^X,i^X)$ and $(i^X+N,i^X+N)$.}, i.e. the index where lead $X$ is
attached. The positive dimensionless real constant $\eta^X$ determines the
strength of the lead's coupling to the system, and $\mu^X$ is the spin
accumulation---i.e.  the difference in chemical potential between spin-up and
spin-down electrons---in the lead, generated, for example, by the spin Hall
effect. In this work, we attach at most two leads: the left lead
($X=\mathrm{L}$) at $i^\mathrm{L}=1$ and optionally the right lead
($X=\mathrm{R}$) at $i^\mathrm{R}=N$. We choose the coupling for positive and
negative modes to be equal-but-opposite [indicated by $\mu^X\sigma_3$ in
Eq.~(\ref{eq:lead-se})], such that our system reduces to the one considered by
\citet{2017PhRvB..96q4422Z} in the limit $K\to 0$ (up to the splitting into
positive and negative modes itself, which, at $K=0$, becomes a purely
notational operation). At the level of the approximations used by
\citet{2017PhRvB..96q4422Z}, the lead self-energy for this geometry is
determined only by the electrons in the metal and the interfacial interaction,
and is independent of the magnons and their particle-hole structure, making
the form of Eq.~(\ref{eq:lead-se}) a natural choice for our model.

The form of the lead self-energy given by Eq.~(\ref{eq:lead-se}) is only valid
when one assumes the spin basis is the natural basis for the lead
Hamiltonians, i.e., that the leads inject a well-defined amount of spin into
the ferromagnet. This is the case provided the electron spin in the leads is
polarized in the $z$-direction and a spin-flip scattering process at the
interface is the source of magnons: here a spin-$\frac{1}{2}$ excitation in
the leads is flipped to $-\frac{1}{2}$, injecting a (spin-1) HP magnon into
the ferromagnet. In the presence of anisotropy, the circular HP magnon is a
superposition of elliptical magnons.

To find an expression for the Gilbert-like damping self-energy, it is
important to carefully consider what one would expect the state of the system
to be in thermal equilibrium. Given that the lead contributions are local,
acting on only one or two sites of a much larger bulk, we assume our system
ultimately thermalizes to states close to the eigenstates of the free
anisotropic ferromagnet, i.e. the elliptical quasiparticles. Thus, what is
linearly damped in our system is the density of elliptical magnons, which does
not necessarily correspond to the classical magnetization---hence our use of
the term `Gilbert\emph{-like} damping', as opposed to just `Gilbert damping':
the latter, in the strict sense, refers to damping of the classical
magnetization only \cite{2004ITM....40.3443G}.

By this rationale, we employ a simple linear damping self-energy \emph{in the
elliptical basis}:\begin{align}
    \Sigma^\mathrm{B,ell}&=-\upi\alpha\omega.
\end{align}
Here $\mathrm{B}$ stands for `bulk' (as this is the only bulk self-energy we
take into account), and $\alpha$ is the Gilbert-like damping
parameter.

Transforming to the spin basis, we find\begin{align}
    \Sigma^\mathrm{B}(\omega)&=-\upi\alpha\omega
            \mathcal{T}^\dagger\mathcal{T},
\end{align}
where $\mathcal{T}^\dagger\mathcal{T}$ becomes the identity matrix in the
limit $K\to0$. In this limit, the bulk self-energy reduces to standard Gilbert
damping, which has been addressed by \citet{2017PhRvB..96q4422Z}.

The total (retarded) self-energy in our model is then simply the sum of the
lead- and bulk self-energies in the spin basis:\begin{align}
    \Sigma(\omega)&=\Sigma^\mathrm{B}(\omega)+\Sigma^\mathrm{L}(\omega)
            +\Sigma^\mathrm{R}(\omega).
\end{align}

Under the assumption that the lead and bulk thermal baths are sufficiently
large to be undisturbed by coupling to the spins, we may use the
fluctuation-dissipation theorem \cite{rammer2007quantum} to find the
associated Keldysh self-energy:\begin{align}
    \Sigma^\mathrm{K}(\omega)
        &=2\Sigma^\mathrm{B}(\omega)\mathcal{T}^{-1}
            F^\mathrm{B}(\omega)\mathcal{T} \nonumber \\
        &\qquad+2\Sigma^\mathrm{L}(\omega)F^\mathrm{L}(\omega) 
        +2\Sigma^\mathrm{R}(\omega)F^\mathrm{R}(\omega).
    \label{eq:sigmaK}
    \intertext{Here, we define the statistical matrix}
    F^X(\omega)&\equiv\diag\Bigg\{
            \coth\left(\frac{\omega-\mu^X}{2\kBT^X}\right), \nonumber \\
        &\hspace{6em}-\coth\left(\frac{-\omega-\mu^X}{2\kBT^X}\right)
            \Bigg\}, \label{eq:statmat}
\end{align}
with $X\in\{\mathrm{B},\mathrm{L},\mathrm{R}\}$, $\kB$ the Boltzmann constant,
and $T^X$ the temperature of the subsystem $X$. We will further assume the
magnon chemical potential vanishes ($\mu^\mathrm{B}=0$), such that
$\mathcal{T}^{-1}F^\mathrm{B}(\omega)\mathcal{T}
=\coth\left(\frac{\omega}{2\kBT^\mathrm{B}}\right)$ is a real number
multiplying the identity matrix.

Finally, from the Keldysh self-energy, we compute the Keldysh Green's
function \cite{keldysh1965diagram,haug2008quantum}\begin{align}
    g^\mathrm{K}(\omega)&=g(\omega)\Sigma^\mathrm{K}(\omega)g^\dagger(\omega).
\end{align}
Note that $g^\mathrm{K}(\omega)$ is symmetric and anti-hermitian, and
therefore pure-imaginary.

\subsection{Observables}
\label{subsec:observ}
Using the elements outlined in Section~\ref{subsec:framework}, we may compute
any physical observable of our system. As we are primarily interested in
steady-state behavior, the most obvious objects to consider are the equal-time
two-point functions of the creation and annihilation operators of HP magnons.
In the presence of anisotropy, we expect to obtain nonzero anomalous
correlations, e.g. $\big<b_ib_j\big>$, because the states in the system are a
superposition of HP magnon states (leading to nonconservation of spin).  The
normal and anomalous correlation functions are conveniently collected in a
single matrix through the vector operator $\phi$, e.g.\begin{align}
    \upi g^>_{ij}(t)&=\left<\phi_i(t)\phi^\dagger_j(t)\right>=
            \left<\begin{pmatrix}b_i(t)\\b^\dagger_i(t)\end{pmatrix}\otimes
            \begin{pmatrix}b^\dagger_j(t)&b_j(t)\end{pmatrix}\right>
            \nonumber \\
        &=\begin{pmatrix}
                \Big<b_i(t)b^\dagger_j(t)\Big>
                &\Big<b_i(t)b_j(t)\Big> \\
                \Big<b^\dagger_i(t)b^\dagger_j(t)\Big>
                &\Big<b^\dagger_i(t)b_j(t)\Big>
            \end{pmatrix}.
\end{align}

Conversely, we may compute two-point functions of the elliptical magnons
$\Psi\equiv\mathcal{T}\phi\equiv\begin{pmatrix}\psi,&\psi^\dagger\end{pmatrix}$,
e.g.\begin{align}
    \left<\Psi^\dagger(t)\Psi(t)\right>
        &=\mathcal{T}^*\left<\phi^\dagger(t)\phi(t)\right>
            \mathcal{T}^\mathrm{T}.
\end{align}
Here, we expect the anomalous blocks to be nonzero only when lead coupling and
anisotropy are simultaneously present: if only anisotropy is present, there
are no damping terms that try to push the system away from the native
elliptical magnon eigenstates (spin is not conserved, but there are no
explicit sources and sinks of spin). Conversely, if lead coupling is present
but anisotropy is absent, the elliptical magnons are identical to the HP ones,
there is no breaking of spin conservation, and the system reduces to the case
investigated by \citet{2017PhRvB..96q4422Z}.

As stated in Section~\ref{subsec:framework}, the matrix
$\rho_{\mu\nu}=\big<\phi^\dagger_\mu\phi_\nu\big>$ (at some arbitrary time in
the steady state, and with the indices $\mu$ and $\nu$ in the range $[1,2N]$)
containing number densities and off-diagonal correlations may be computed
through the lesser Green's function $g^<$:\begin{align}
    \label{eq:rho}
    \rho=\upi g^<&=\upi\int\frac{\dif\omega}{2\uppi}
            \,g^<(\omega) \\
        &=\frac{\upi}{2}\int\frac{\dif\omega}{2\uppi}\,
            \left(g^\mathrm{K}(\omega)-g(\omega)
            +g^\dagger(\omega)\right). \nonumber
\end{align}
For the sake of brevity, we shall refer to $\rho$ as the density matrix,
although the off-diagonal components are in fact off-diagonal correlations.
Note also that by the symmetry of $g$ and $g^\mathrm{K}$, and anti-hermiticity
of the Keldysh Green's function, the lesser Green's function is itself
symmetric, anti-hermitian and pure-imaginary. One may alternatively work
directly with the Keldysh Green's function, of which the corresponding
observable is the semiclassical (SC) HP magnon density matrix\begin{align}
    \rho^\mathrm{SC}=\frac{\upi}{2} g^\mathrm{K}-\frac{1}{2}
        &=\frac{1}{2}\left<\left\{\phi^\dagger,\phi\right\}\right>
            -\frac{1}{2} \nonumber \\
        &=\frac{\upi}{2}\int\frac{\dif\omega}{2\uppi}\,
            g^\mathrm{K}(\omega)-\frac{1}{2}.
\end{align}
In equilibrium, the top-left and off-diagonal blocks correspond directly to
those of the true density matrix of Eq.~(\ref{eq:rho}).

From the density matrix $\rho$, we may compute the uncertainty operators
$\Delta \bar{S}^x$ and $\Delta \bar{S}^y$ for the corresponding normalized
spin operators $\bar{S}^x$ and $\bar{S}^y$, which allow us to determine
whether the elliptical magnons are squeezed in phase space
\cite{1983Natur.306..141W}. From the normalized spin
operators\begin{subequations}\label{eq:hptransnorm}\begin{align}
    \bar{S}^x_i&=\frac{1}{\sqrt{2}}\left(b_i+b^\dagger_i\right),\\
    \bar{S}^y_i&=\frac{-\upi}{\sqrt{2}}\left(b_i-b^\dagger_i\right),
    \intertext{and}
    \bar{S}^z_i&=1-b^\dagger_ib_i,
\end{align}\end{subequations} [i.e. the HP transformation with $S$ set to 1,
and applied in reverse with respect to Eqs.~(\ref{eq:hptrans})], we
immediately find the uncertainty operators\begin{subequations}
\label{eq:uncert}
\begin{align}
    \Delta \bar{S}^x_i&\equiv\sqrt{\big<(\bar{S}^x_i)^2\big>
            -\big<\bar{S}^x_i\big>^2} \nonumber \\
        &=\sqrt{\frac{1}{2}\left[\big<b_ib^\dagger_i\big>
            +\big<b^\dagger_ib_i\big>+\big<b_ib_i\big>
            +\big<b^\dagger_ib^\dagger_i\big>\right]} \nonumber \\
        &=\sqrt{\frac{1}{2}\left[I_+\rho I_+^\mathrm{T}\right]_{ii}}
    \intertext{and}
    \Delta \bar{S}^y_i&\equiv\sqrt{\big<(\bar{S}^y_i)^2\big>
            -\big<\bar{S}^y_i\big>^2} \nonumber \\
        &=\sqrt{\frac{1}{2}\left[\big<b_ib^\dagger_i\big>
            +\big<b^\dagger_ib_i\big>-\big<b_ib_i\big>
            -\big<b^\dagger_ib^\dagger_i\big>\right]} \nonumber \\
        &=\sqrt{\frac{1}{2}\left[I_-\rho I_-^\mathrm{T}\right]_{ii}},
    \intertext{where $I_\pm$ are the $N\times 2N$ matrices}
    I_\pm&\equiv\delta_{ij}(1, \pm1).
\end{align}\end{subequations}
Here, the one-point functions $\big<\bar{S}^x_i\big>$ and
$\big<\bar{S}^y_i\big>$ vanish, because we do not explicitly couple to a
pumping field and are not considering Bose-Einstein condensates
\cite{2012PhRvL.108x6601B}. The Robertson uncertainty principle
\cite{1929PhRv...34..163R} then states that $\Delta \bar{S}^x_i\Delta
\bar{S}^y_i\ge\frac{1}{2}$. If either $\Delta \bar{S}^x_i<\frac{1}{\sqrt{2}}$
or $\Delta \bar{S}^y_i<\frac{1}{\sqrt{2}}$, the state is squeezed
\cite{1983Natur.306..141W}, and the pattern of quantum fluctuations of the
spin around the $z$-axis takes the form of an ellipse, rather than a circle
\cite{2020ApPhL.117i0501K}. As noted by \citet{2020ApPhL.117i0501K}, the
purely quantum mechanical squeezing should not be confused with the
magnetization trajectory of a classical elliptical spin wave: the latter
concerns coherent excited states, whereas squeezing persists even in the
ground state and affects properties such as entanglement.

In addition to the magnon density and the related observables, we may compute
the spin currents in our system. These follows from the continuity equation of
the magnetization; a brief outline of the derivation is given in
Appendix~\ref{app:spincur}. The total spin current
$j_\mathrm{s,tot}^\mathrm{L}$ flowing out of the left lead comprises three
Landauer-Büttiker-type \cite{ventra2008electrical} terms:\begin{align}
j_\mathrm{s,tot}^\mathrm{L}(t)&=j_\mathrm{s}^{\mathrm{R}\to\mathrm{L}}
                +j_\mathrm{s}^{\mathrm{B}\to\mathrm{L}}
                +j_\mathrm{s}^{\mathrm{L}}
    \intertext{where}
    j_\mathrm{s}^X&=-\real\Tr\int\frac{\dif\omega}{2\uppi}\iota^{X}(\omega).
\end{align}
Here the integrands $\iota^X$ are the tunneling term\begin{subequations}
\label{eq:iotas}
\begin{align}
    \label{eq:iotaR}
    \iota^{\mathrm{R}\to\mathrm{L}}(\omega)
        &=g^\dagger(\omega)\sigma_3\Sigma^\mathrm{L}(\omega)g(\omega)
            \Sigma^\mathrm{R}(\omega) \nonumber \\
        &\hspace{5em}\times\left[F^\mathrm{R}(\omega)
            -F^\mathrm{L}(\omega)\right],
    \intertext{the bulk term}
    \iota^{\mathrm{B}\to\mathrm{L}}(\omega)
        &=g^\dagger(\omega)\sigma_3\Sigma^\mathrm{L}(\omega)g(\omega)
            \Sigma^\mathrm{B}(\omega) \nonumber \\
        &\hspace{5em}\times\left[\mathcal{T}^{-1}F^\mathrm{B}(\omega)
            \mathcal{T}-F^\mathrm{L}(\omega)\right],
    \label{eq:iotaB}
    \intertext{and the lead-local term}
    \label{eq:iotaL}
    \iota^{\mathrm{L}}(\omega)
        &=g^\dagger(\omega)\sigma_3\Sigma^\mathrm{L}(\omega)g(\omega)h
            \nonumber \\
        &\hspace{5em}\times\left[\left.F^\mathrm{L}(\omega)
            \right|_{\mu^\mathrm{L}=0}-F^\mathrm{L}(\omega)\right].
\end{align}
\end{subequations}

Conversely, the spin current out of the right lead consists of the same
expressions but with $\mathrm{L}$ and $\mathrm{R}$ swapped.

Note that the terms in $j_\mathrm{s,tot}^\mathrm{L}$ contain the statistical
matrices $F^X(\omega)$ in place of the scalar Bose-Einstein functions one
would normally find in Landauer-Büttiker equations. One may exploit various
symmetries of the components of the integrals to recover the more familiar
form (in the circular limit identical to the expressions given by
\citet{2017PhRvB..96q4422Z}), though this requires the transmission functions
to be written in terms of the individual $N\times N$ blocks of the component
$2N\times 2N$ matrices, as has been done by e.g. \citet{2020AnPhy.41268010R}.

\begin{figure}
    \subfloat[][]{
        \label{fig:resnetwork-full}
        \includegraphics[width=\linewidth]{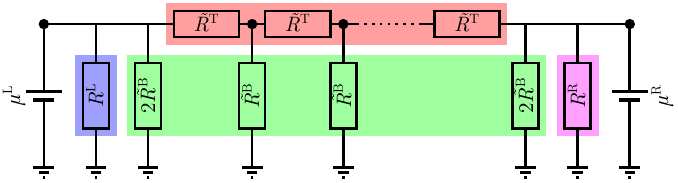}
    }

    \subfloat[][]{
        \label{fig:resnetwork-reduced}
        \includegraphics[width=\linewidth]{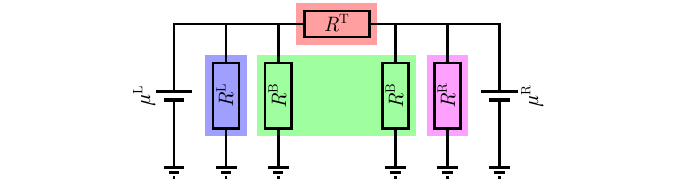}
    }
    \caption[]{Resistor network diagrams of our model system.
    \subref{fig:resnetwork-full}: In the full diagram, each black dot represents a
    spin, each resistor $\tilde{R}^\mathrm{T}$ represents an inter-spin
    coupling, and each resistor $\tilde{R}^\mathrm{B}$ represents a coupling
    to the field(s) responsible for Gilbert-like damping. The spin currents
    through the resistors $R^\mathrm{L/R}$ are given by the lead-local terms
    $-j^\mathrm{L/R}$. \subref{fig:resnetwork-reduced}: By repeated
    application of the $\Delta-Y$ transform \cite{kennelly1899equivalence},
    one may collapse the internal couplings (red and green blocks), thereby
    reducing the $N$-spin resistor network to a five-resistor form. In this
    reduced diagram, the current through $R^\mathrm{T}$ (from right to left)
    is given by $j^\mathrm{R\to L}$ and the currents through the left (right)
    resistors $R^\mathrm{B}$ are represented by the bulk terms
    $-j^\mathrm{B\to L(R)}$. The lead-local resistors remain unchanged.}
    \label{fig:resnetwork}
\end{figure}

To interpret the three spin current contributions, it is useful to consider
the system as a spin resistor network, shown in
Fig.~\ref{fig:resnetwork-full}. Here each node in the circuit represents a
spin in our chain, and each resistor represents a coupling either between
spins (tunneling resistors $\tilde{R}^\mathrm{T}$) or to a damping element
(Gilbert-like damping for the bulk resistors $\tilde{R}^\mathrm{B}$, and lead
damping for $R^\mathrm{L/R}$).  The system is biased at either lead with a
spin accumulation (voltage) $\mu^\mathrm{L/R}$. In this resistor network
analogy, one may `integrate out' the bulk spins by repeated application of the
$\Delta-Y$ transform \cite{kennelly1899equivalence} to obtain
Fig.~\ref{fig:resnetwork-reduced}. At equal temperature
($T^\mathrm{L}=T^\mathrm{R}=T^\mathrm{B}=T$), the spin currents $j^\mathrm{X}$
may then be interpreted as follows:

The tunneling term $j^\mathrm{R\to L}$ is the current flowing from right to
left through the resistor $R^\mathrm{T}$ in
Fig.~\ref{fig:resnetwork-reduced}, and corresponds to the spin current
flowing out of the left lead when a spin accumulation is applied at the right
lead. Physically, it is the term corresponding to magnon-mediated non-local
transport, and roughly corresponds to the current measured experimentally in a
ferromagnetic insulator by \citet{2015NatPh..11.1022C}, although our work
considers the ballistic regime ($\mu^\mathrm{B}=0$) rather than the diffusive
regime.

The bulk term $j^\mathrm{B\to L}$ corresponds to the current flowing out
of the left lead as a result of Gilbert-like damping in the bulk. It is
negative when a positive spin accumulation is applied to the left lead,
indicating spin current flows from the lead into the bulk, where it is
dissipated into the lattice. In Fig.~\ref{fig:resnetwork-reduced},
$-j^\mathrm{B\to L(R)}$ is the current flowing to ground through the left
(right) resistor $R^\mathrm{B}$.

The lead-local term $j^\mathrm{L}$, corresponding to the current flowing
from ground upwards through $R^\mathrm{L}$ in
Fig.~\ref{fig:resnetwork-reduced}, is unique to systems that exhibit the
Bogoliubov structure described at the start of this section. It is linear in
$K$ to lowest nonvanishing order, and, at nonzero $K$, vanishes unless the
system is driven by the application of an electronic spin accumulation in the
lead. We may therefore conclude that it arises due to the mismatch between the
lead states, where spin is a good quantum number, and the elliptical magnon
eigenstates of the anisotropic ferromagnet. Ultimately, the mismatch is
necessarily compensated by the lattice \cite{2020PhRvB.101i4402Z}. As this
term contributes directly to the spin current flowing out of the lead to which
a spin bias is applied, it offers a way to probe the ellipticity of magnons
through local spin current measurements.

Taking the resistor network analogy further, the reduced model of
Fig.~\ref{fig:resnetwork-reduced} provides us with a new set observables
more generic than the spin currents themselves, namely the spin resistances
$R^\mathrm{T}$, $R^\mathrm{L/R}$ and $R^\mathrm{B}$. Setting
$\mu^\mathrm{R}=0$ and formally expanding the left-lead spin current terms in
$\mu^\mathrm{L}$, we obtain\begin{subequations}\label{eq:lincur}\begin{align}
    j_\mathrm{s}^\mathrm{R\to L}
        &=j_\mathrm{s0}^\mathrm{R\to L}(T^\mathrm{L}, T^\mathrm{R})
            -\frac{1}{R^\mathrm{T}}\mu^\mathrm{L}, \\
    j_\mathrm{s}^\mathrm{B\to L}
        &=j_\mathrm{s0}^\mathrm{B\to L}(T^\mathrm{L}, T^\mathrm{B})
            -\frac{1}{R^\mathrm{B}}\mu^\mathrm{L}, \\
    j_\mathrm{s}^\mathrm{L}&=-\frac{1}{R^\mathrm{L}}\mu^\mathrm{L}.
\end{align}\end{subequations}
Here $j_\mathrm{s0}^\mathrm{R\to L}(T^\mathrm{L}, T^\mathrm{R})$ and
$=j_\mathrm{s0}^\mathrm{B\to L}(T^\mathrm{L}, T^\mathrm{B})$ are spin Seebeck
effect \cite{2012NatMa..11..391B,2020AIPA...10a5031O} terms that vanish when
$T^\mathrm{L}=T^\mathrm{R}$ and $T^\mathrm{L}=T^\mathrm{B}$, respectively.

\section{Numerical implementation and results}
\label{sec:results}
The framework outlined in the previous section is implemented numerically for
system sizes of order $N=20$. At low or moderate damping, the functions in our
setup are sharply peaked in the frequency domain; frequency integrals are
evaluated with an adaptive trapezoidal algorithm to avoid missing such peaks.
As the setup requires matrices of size $2N\times2N$ and the computation of
observables includes one matrix inversion and multiple dense matrix
multiplications per frequency sample, the numerical implementation scales
poorly with system size. However, as the qualitative differences between
systems of size $N=40$ and $N=20$ turn out to be minimal, we believe the
latter to be a fair compromise between manageable computation time and
sufficient capture of large-system behavior.

Our use of simplistic linear damping leads to a logarithmic divergence if the
frequency integrals in the expressions for $\rho$ or $\rho^\mathrm{SC}$ are
taken from $-\infty$ to $\infty$. We regularize the integrals by restricting
the integration interval to $[-\varepsilon_\mathrm{max},
\varepsilon_\mathrm{max}]$, where\begin{align}
    \varepsilon_\mathrm{max}&=\lim_{N\to\infty}\varepsilon_N.
\end{align}

We seek to investigate qualitative changes in the behavior of our system as
the anisotropy $K$ is increased, while mitigating the effects of changes to
the energetics of the ferromagnet's eigenstates. To realize this, we shall
keep the elliptical magnon gap $\varepsilon_1$, given by
Eq.~(\ref{eq:dispersion}), fixed. Furthermore, we keep the exchange-like
constant $J$ fixed, and adjust the field-like parameter
$\Delta=2J\cos\left(\frac{\uppi}{N+1}\right)+\sqrt{\varepsilon_1^2+K^2}$
accordingly. Finally, we shall measure all energy scales relative to $J$,
which is numerically realized by setting $J=1$.

\subsection{Spin conductances}
We compute the spin conductances\begin{subequations}\begin{align}
    G^\mathrm{T}&\equiv\frac{1}{R^\mathrm{T}}, \\
    G^\mathrm{B}&\equiv\frac{1}{R^\mathrm{B}},
    \intertext{and}
    G^\mathrm{L}&\equiv\frac{1}{R^\mathrm{L}}
\end{align}
\end{subequations}
by fitting the components of $j_\mathrm{s,tot}^\mathrm{L}$ to
Eqs.~(\ref{eq:lincur}) for small values of $\mu^\mathrm{L}$, setting
$T^\mathrm{L}=T^\mathrm{R}=T^\mathrm{B}=T$ and $\mu^\mathrm{R}=0$. We consider
a system with parameters $N=20$, $\alpha=0.001$, $\varepsilon_1=0.025J$ and
$\eta^\mathrm{L}=\eta^\mathrm{R}=8$. (Here, the values for $\eta^\mathrm{L/R}$
are chosen in line with \citet{2017PhRvB..96q4422Z}, while our choices for
$\alpha$ and $\varepsilon_1$ are fairly arbitrary within the low-damping and
low-gap regimes, respectively.) Note that because we have set $\hbar=1$,
the conductances are dimensionless.

\begin{figure}
	\centering
	\includegraphics[width=\linewidth]{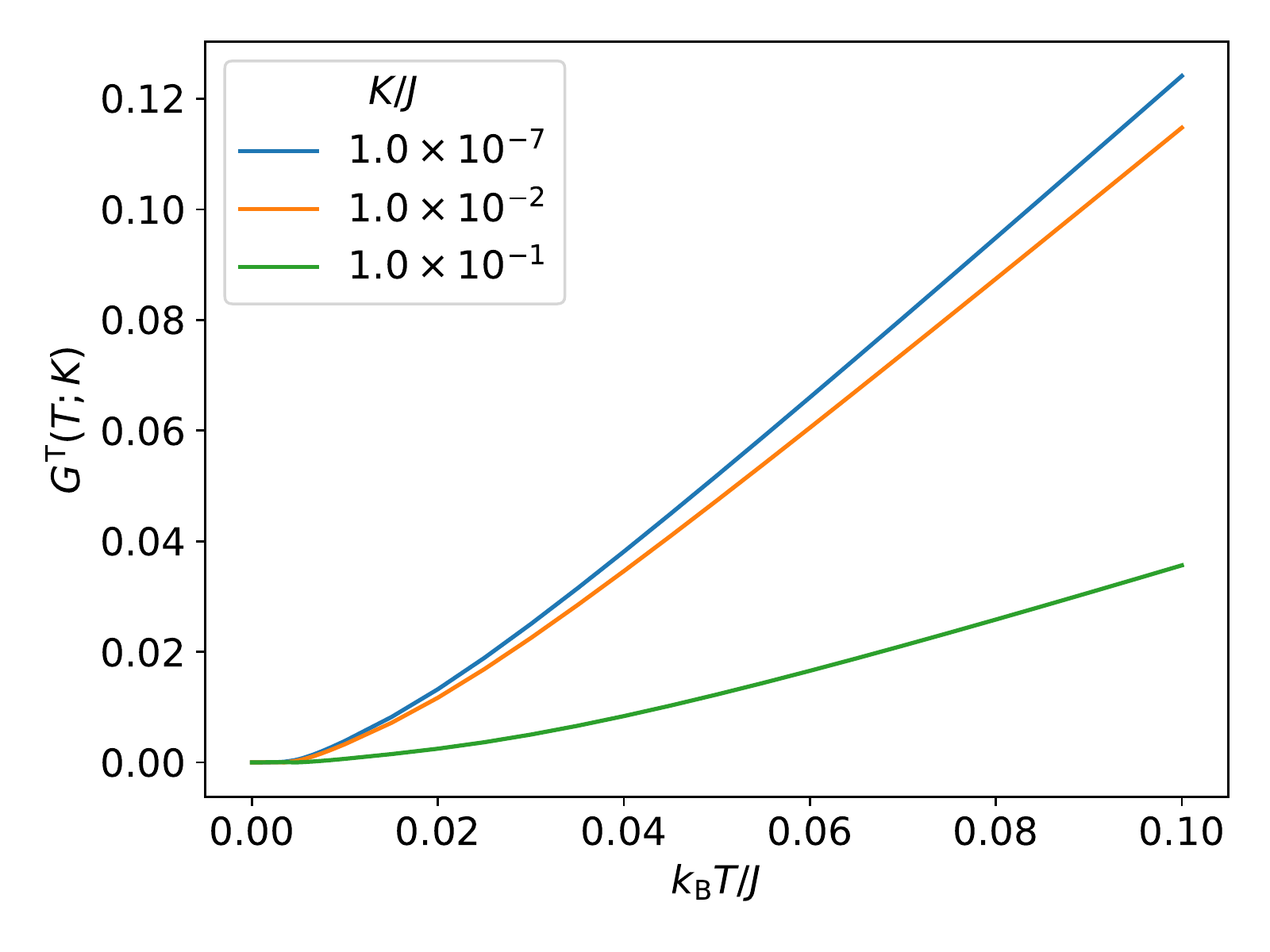}
	\caption{Tunneling conductance versus temperature at magnon gap
    $\varepsilon_1/J=0.025$, for different values of the anisotropy $K$.}
    \label{fig:Gtunnel}
\end{figure}

Figure~\ref{fig:Gtunnel} shows the tunneling conductance $G^\mathrm{T}$ vs.
temperature $\kBT$ at various values of $K$. In all cases, the tunneling
conductance vanishes at $T=0$ (to numerical accuracy; the highly nonlinear
behavior at low temperature limits the fitting accuracy) and slowly
transitions to being linear with temperature.  The effect of anisotropy is to
suppress the conductance, although this effect is small until
$K/J=\mathcal{O}(0.1)$, i.e. very large anisotropy (e.g. for yttrium-iron
garnet, a comparison of literature values
\cite{2020JMMM..51467099S,2017PhRvB..95a4423X,edmonds1959effective,1993PhR...229...81C,2017npjQM...2...63P}
yields $K/J=\mathcal{O}(10^{-3}\text{---}10^{-2})$, although the range is
highly variable between different materials \cite{coey2009magnetism}).
Physically, this may be understood by the fact that the leads are not
commensurate to the elliptical spin waves, which causes an increase in
reflection at the interface.

\begin{figure}
	\centering
	\includegraphics[width=\linewidth]{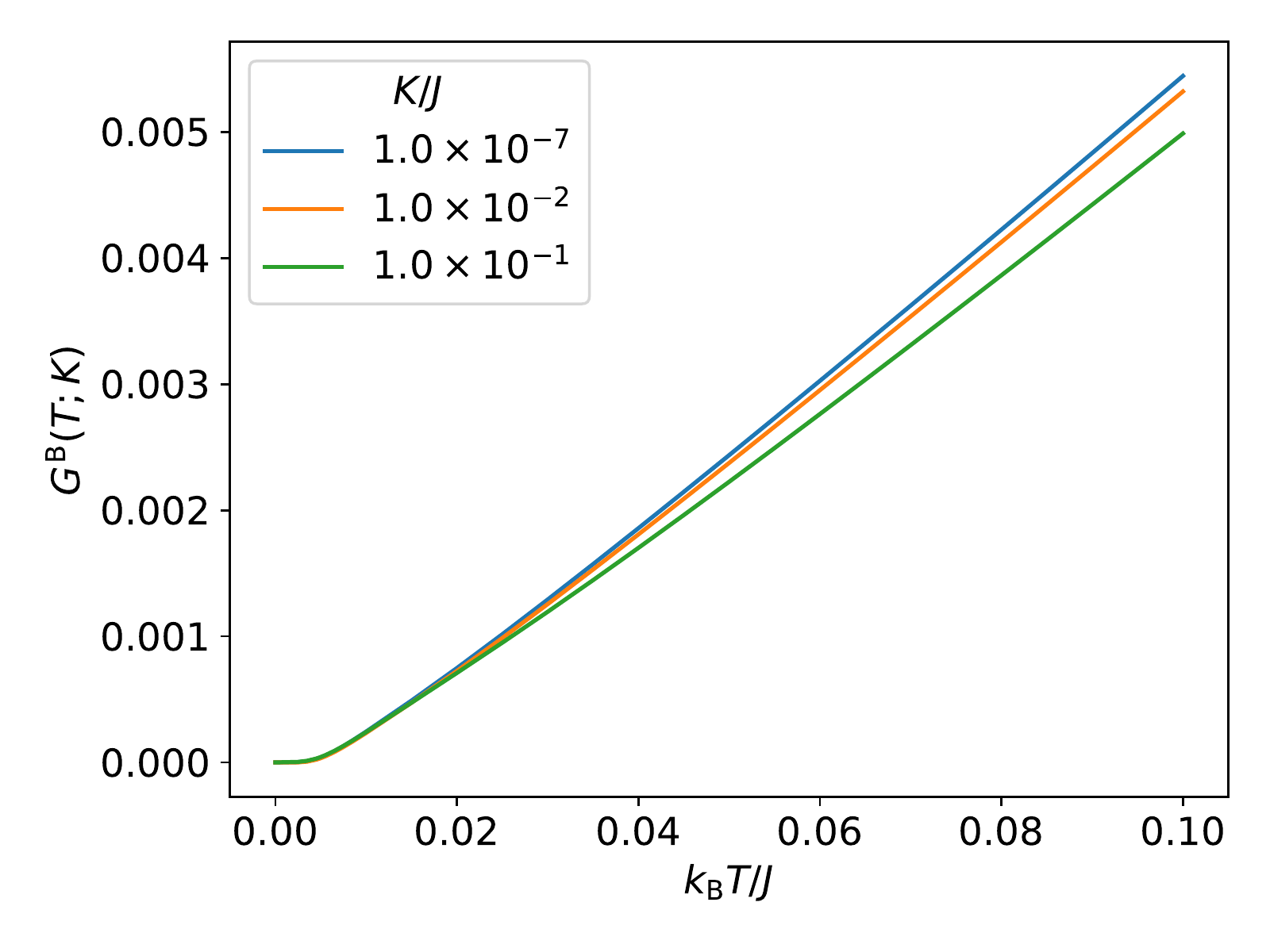}
	\caption{Bulk conductance versus temperature at magnon gap
    $\varepsilon_1/J=0.025$, for different values of the anisotropy $K$.}
    \label{fig:Gbulk}
\end{figure}

The bulk conductance $G^\mathrm{B}$, shown in Fig.~\ref{fig:Gbulk}, similarly
vanishes at $T=0$ and is suppressed by anisotropy. Unlike the tunneling
conductance, where the transition to a linear regime is smeared out at higher
anisotropies, the bulk conductance transitions more abruptly, at $\kBT/J\sim
0.005$.  The suppression with anisotropy is mild: at $\kBT/J=0.1$, increasing
the anisotropy from $K/J=\num{1e-7}$ to $K/J=\num{1e-1}$ suppresses the bulk
conductance by roughly 8\%. Our formalism does not elucidate the physical
mechanism underlying this suppression, however, given its small magnitude, we
believe it to be a natural consequence of the anisotropy-dependence of the
dispersion, rather than being the result of any nontrivial effect.

The relatively abrupt transition to a linear regime is a direct consequence of
the low-$\mu^\mathrm{L}$ behavior of the difference of statistical matrices in
the bulk spin current integrand (\ref{eq:iotaB}):\begin{align}
    \coth\left(\frac{\omega}{2\kBT}\right)
        &\mp\coth\left(\frac{\pm\omega-\mu^\mathrm{L}}{2\kBT}\right) \nonumber \\
        &\approx\pm\frac{\mu^\mathrm{L}}{\kBT-\kBT\cosh\left(
            \frac{\omega}{\kBT}\right)}.
\end{align}
Given that the most significant contribution to $G^\mathrm{B}$ arises from a
narrow region of $\iota^\mathrm{B\to L}$ centred around $\omega=\varepsilon_1$
(as one would expect), we may judiciously substitute
$\omega=\varepsilon_1=0.025J$ in this expression. Dividing by
$\mu^\mathrm{L}$, we then obtain a function that exhibits a kink near
$\kBT/J=0.005$, similar to the bulk conductance. We may thus conclude,
qualitatively, that the kink is explained by the requirement for the
temperature to overcome the finite gap.

\begin{figure}
	\centering
	\includegraphics[width=\linewidth]{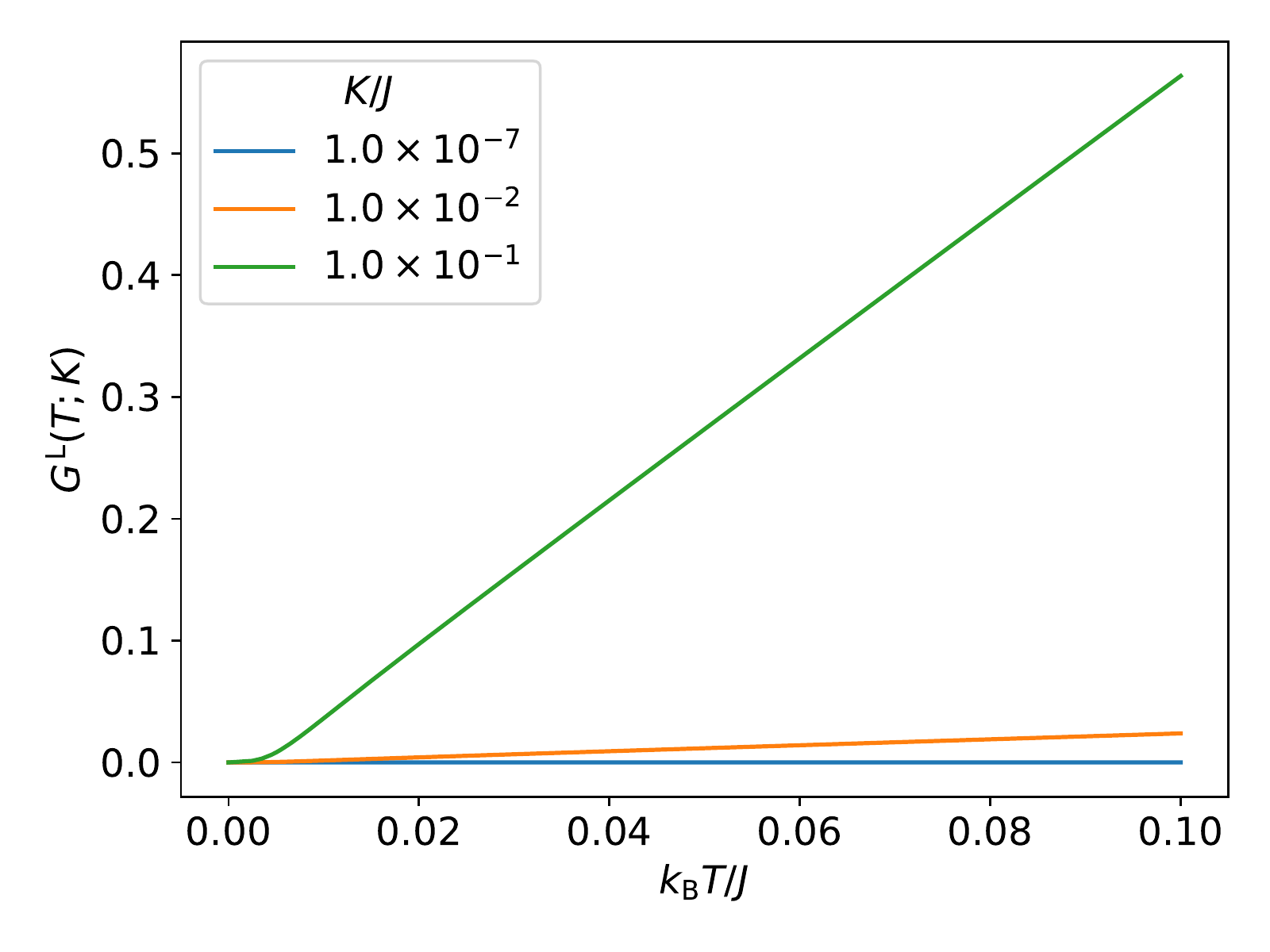}
	\caption{Lead-local conductance versus temperature at magnon gap
    $\varepsilon_1/J=0.025$, for different values of the anisotropy $K$.}
    \label{fig:Gsite}
\end{figure}

In Fig.~\ref{fig:Gsite}, it can be seen that the lead-local conductance
$G^\mathrm{L}$ nearly vanishes at low anisotropy (as expected) and reaches a
magnitude roughly comparable to the bulk conductance at the fairly high
anisotropy value $K/J=\num{1e-2}$. However, $G^\mathrm{L}$ is vastly enhanced
at the very high anisotropy value $K/J=\mathcal{O}(0.1)$, becoming several
times larger than the tunneling conductance, indicating most spin is lost to
the lattice at the left-lead interface. This bears similarity to the
appearance of evanescent spin waves in anisotropic systems
\cite{2020PhRvB.102j4414P}, however, rigorously showing the relation between
these effects requires reconstructing the classical wave picture from our
formalism, which is beyond the scope of this work. Like the bulk conductance,
the transition to a linear regime in the lead-local conductance is relatively
abrupt for the $K/J=\num{1e-1}$ curve.

Although the tunneling and bulk conductances are suppressed by increasing
anisotropy, the corresponding increase in the lead-local conductance is
greater than the decrease in the sum of tunneling and bulk conductances. In
other words, the conductance of the parallel combination of the three spin
resistors $R^\mathrm{T}$, $R^\mathrm{B}$ and $R^\mathrm{L}$ \emph{increases}
with increasing anisotropy, while the individual conductances of
$R^\mathrm{T}$ and $R^\mathrm{B}$ decrease. Thus, although our model does not
provide an obvious way to separate the bulk and lead-local contributions, it
suggests the presence of anisotropy causes the local spin conductance to
increase, while the nonlocal conductance decreases, thereby potentially
providing an experimental way to probe the anisotropy of a ferromagnetic
insulator using spin current measurements.

\subsection{Correlation functions and squeezing}
\begin{figure}
    \centering
    \subfloat[][]{
        \label{fig:rholesser1a}
        \includegraphics[width=\linewidth]{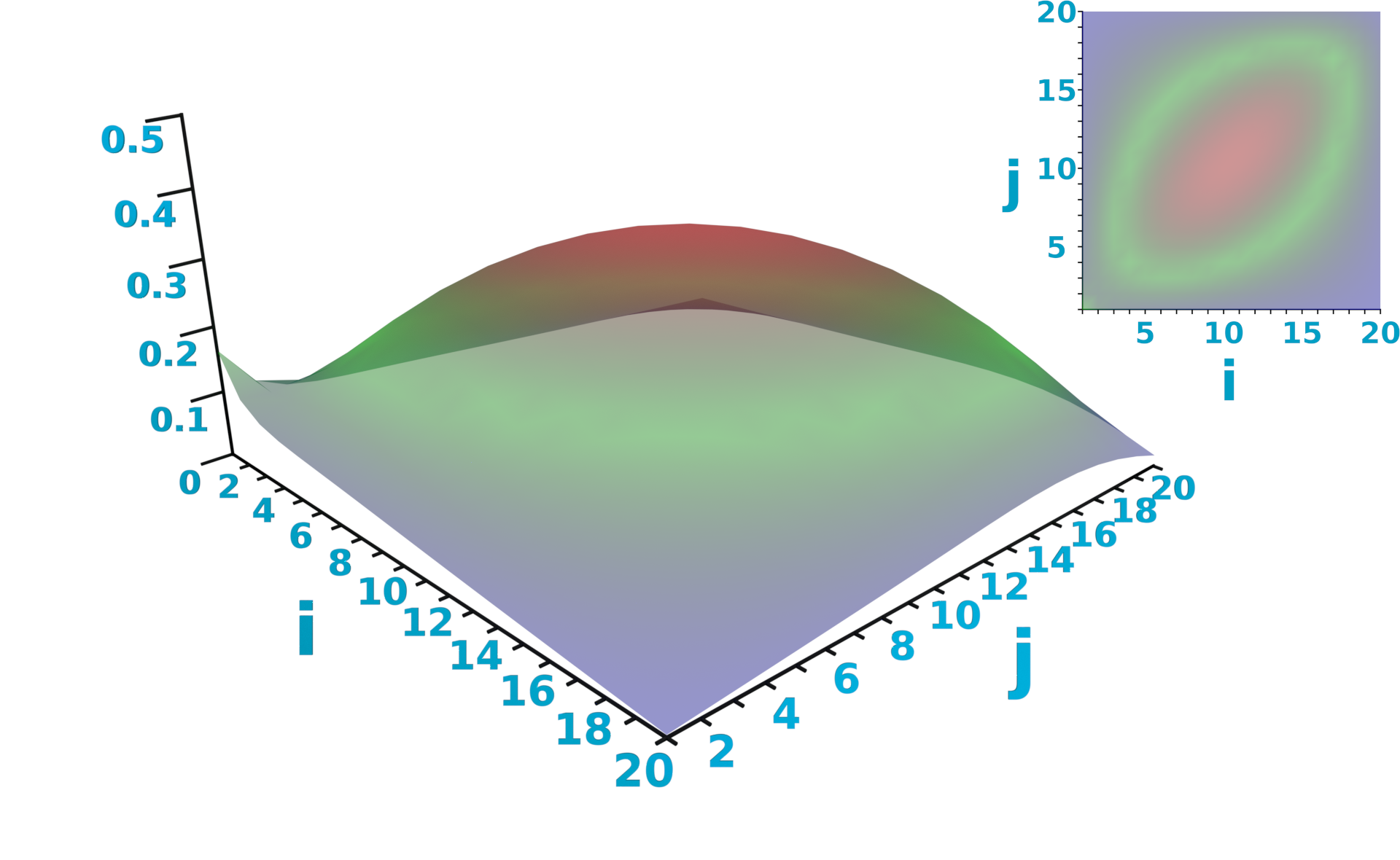}
    }

    \subfloat[][]{
        \label{fig:rholesser1b}
        \includegraphics[width=\linewidth]{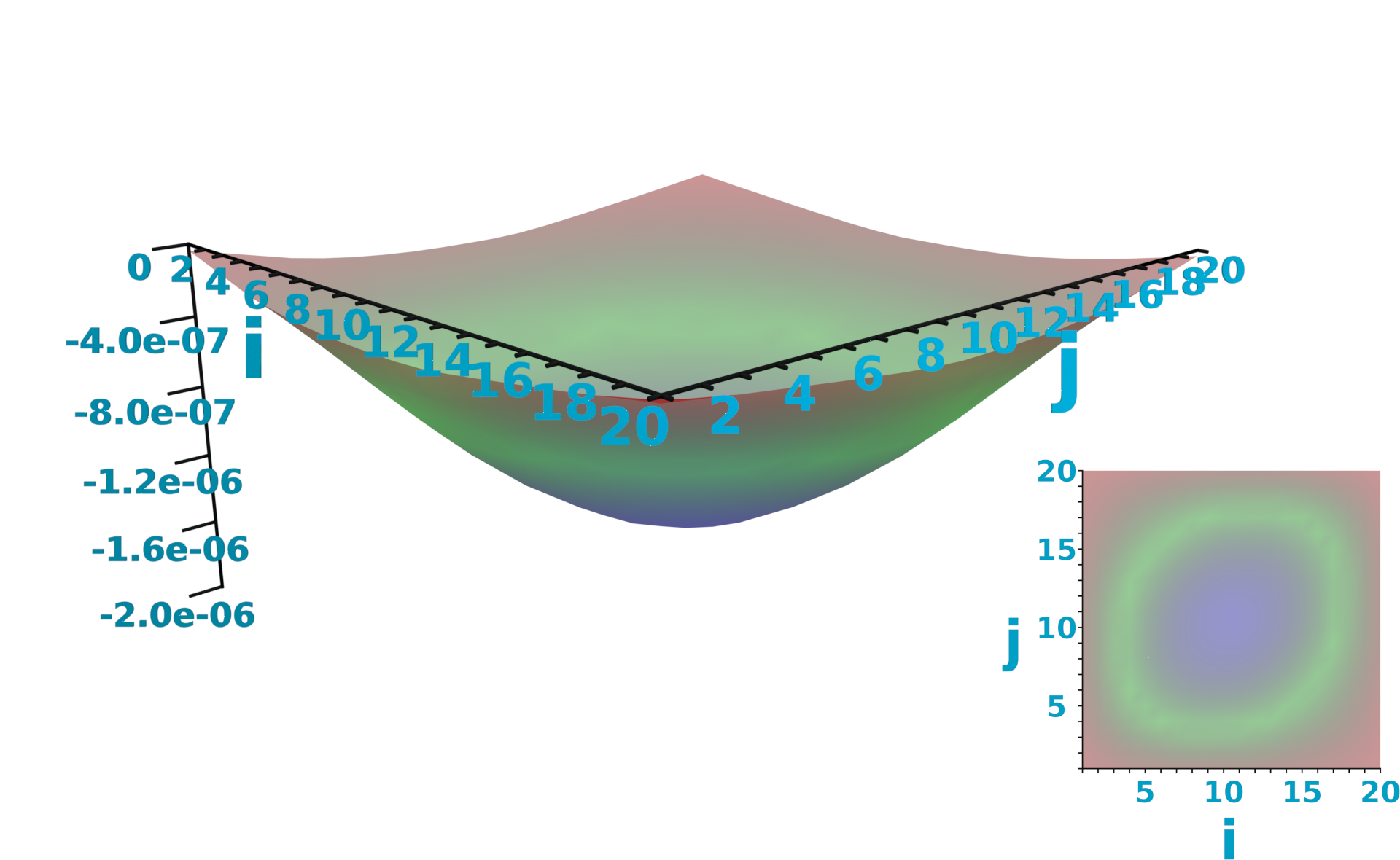}
    }
    \caption[]{Normal and anomalous spin densities for a system of size $N=20$
    with only the left lead attached ($\eta^\mathrm{L}=8$,
    $\eta^\mathrm{R}=0$), low Gilbert-like damping $\alpha=\num{1e-3}$,
    temperature $T/J=0.1$, gap $\varepsilon_1/J=0.025$, and anisotropy
    $K/J=\num{1e-7}$. The horizontal axes represent the site indices $i$ and
    $j$. Insets: heatmaps of the corresponding 3D plots.
    \subref{fig:rholesser1a}: normal density $\big<b^\dagger_i(t)b_j(t)\big>$.
    \subref{fig:rholesser1b}: anomalous density
    $\big<b^\dagger_i(t)b^\dagger_j(t)\big>$.}
    \label{fig:rholesser1}
\end{figure}

To gain insight into the distribution of spin and the profile of spin
nonconservation in the ferromagnet, we compute the density matrix at low and
high anisotropy. Figure~\ref{fig:rholesser1a} shows the spin density matrix
$\big<b^\dagger_i(t)b_j(t)\big>$ in a low-damping ($\alpha=\num{1e-3}$), low
anisotropy ($K/J=\num{1e-7}$) system where only the left lead is attached
($\eta^\mathrm{L}=8$, $\eta^\mathrm{R}=0$) and no biasing is applied
($\mu^\mathrm{L}=\mu^\mathrm{R}=0$). The temperature is taken to be
homogeneous at $\kBT^\mathrm{L}=\kBT^\mathrm{R}=\kBT^\mathrm{B}=0.1J$. The gap
is set to $\varepsilon_1=0.025J$, which is a reasonable value for e.g.
yttrium-iron garnet \cite{1993PhR...229...81C,2014JMMM..356...95K}.

In Fig.~\ref{fig:rholesser1}, the horizontal axes correspond to the site
indices $i$ and $j$. The spin density is slightly elevated at the attached
lead, but primarily accumulates deep within the bulk, taking the shape of
the crest of a standing wave whose wavelength is twice the sample size. Here
it is immediately apparent that the Holstein-Primakoff magnons are
significantly delocalized, as the correlations
$\big<b^\dagger_i(t)b_j(t)\big>$ decrease only slowly as $|i-j|$ grows.

At low anisotropy, the leads and bulk try to drive the system towards the same
set of states, so the anomalous correlations
$\big<b^\dagger_i(t)b^\dagger_j(t)\big>$, shown in Fig.~\ref{fig:rholesser1b},
vanish everywhere up to numerical accuracy. The spin density plots remain
virtually unchanged with increasing anisotropy up until about
$K/J=\mathcal{O}(10^{-3})$.

\begin{figure}
    \centering
    \subfloat[][]{
        \label{fig:rholesser2a}
        \includegraphics[width=\linewidth]{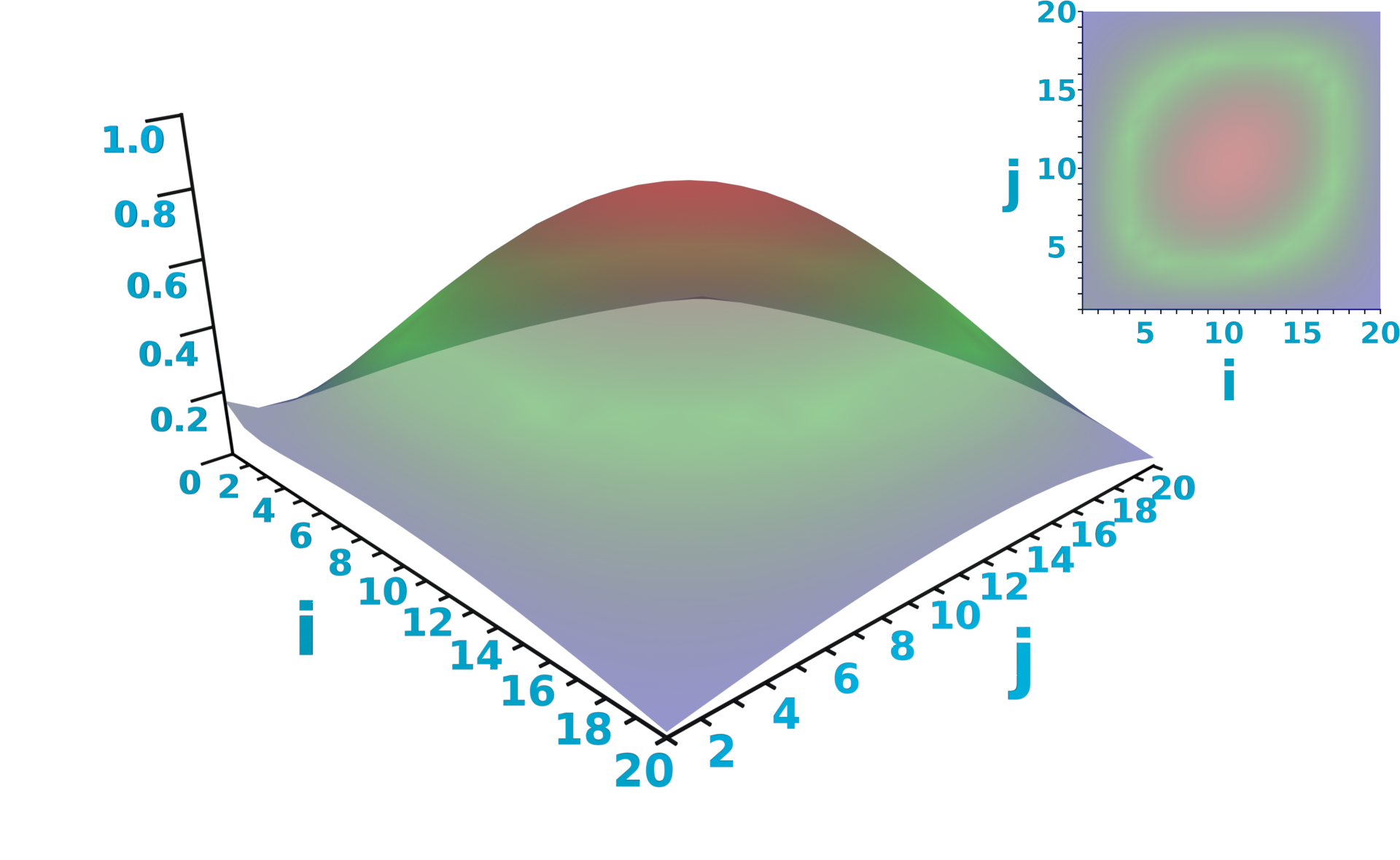}
    }

    \subfloat[][]{
        \label{fig:rholesser2b}
        \includegraphics[width=\linewidth]{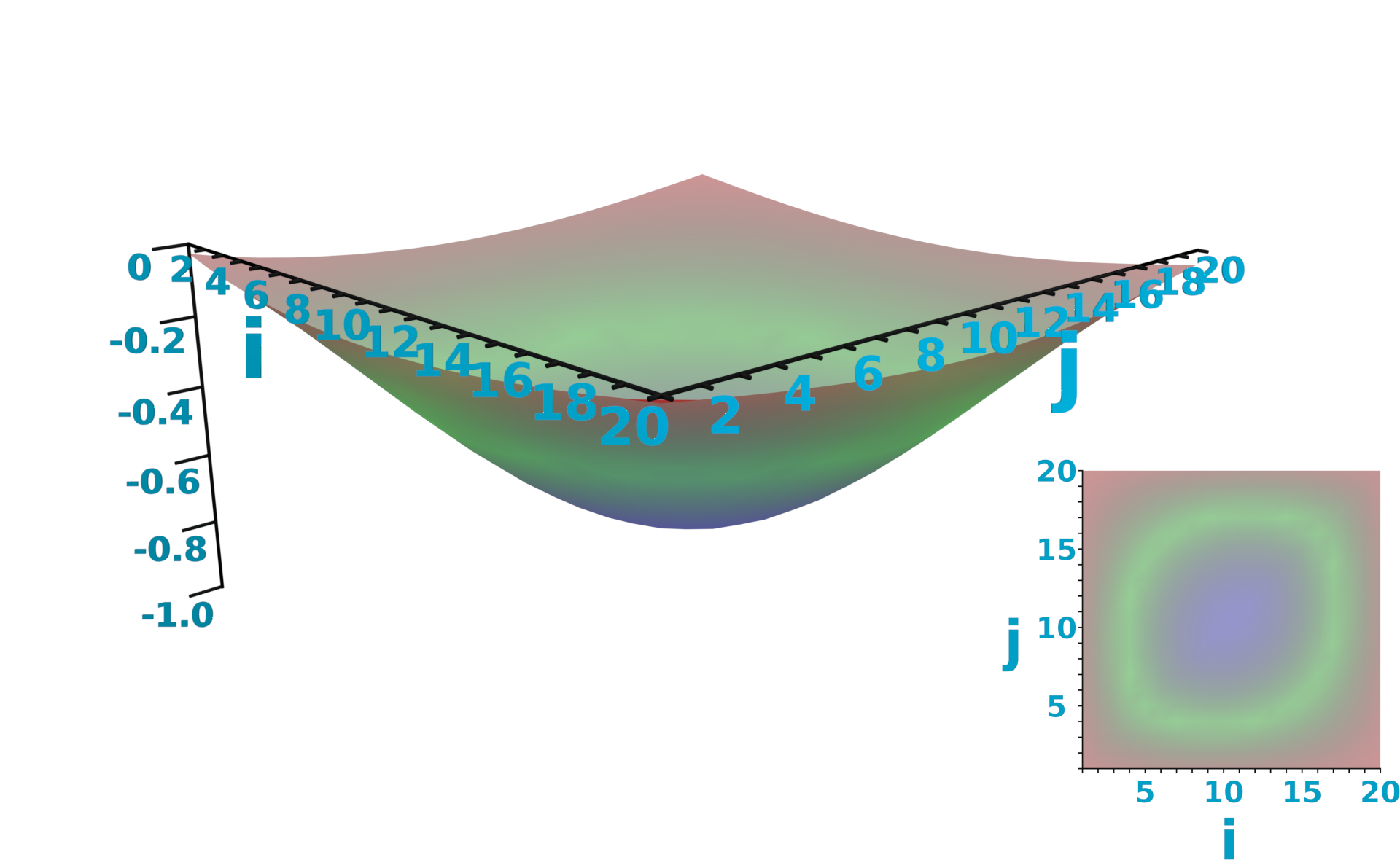}
    }
    \caption[]{Normal and anomalous spin densities for a system with high
    anisotropy, $K/J=\num{5e-2}$. All other parameters are equal to those in
    Fig.~\ref{fig:rholesser1}. \subref{fig:rholesser2a}: normal density
    $\big<b^\dagger_i(t)b_j(t)\big>$. \subref{fig:rholesser2b}: anomalous
    density $\big<b^\dagger_i(t)b^\dagger_j(t)\big>$.}
    \label{fig:rholesser2}
\end{figure}

At much greater anisotropy---$K/J=\num{5e-2}$ shown in
Fig~\ref{fig:rholesser2}---the amplitude of the spin density at the center of
the sample increases significantly (Fig~\ref{fig:rholesser2a}), but the
qualitative appearance of the profile remains broadly the same. However, as
shown in Fig.~\ref{fig:rholesser2b}, the anomalous correlations now take a
large negative value, highlighting that the Holstein-Primakoff magnons are no
longer good basis states in the ferromagnetic bulk.

\begin{figure}
    \centering
    \subfloat[][]{
        \label{fig:rholesser3a}
        \includegraphics[width=\linewidth]{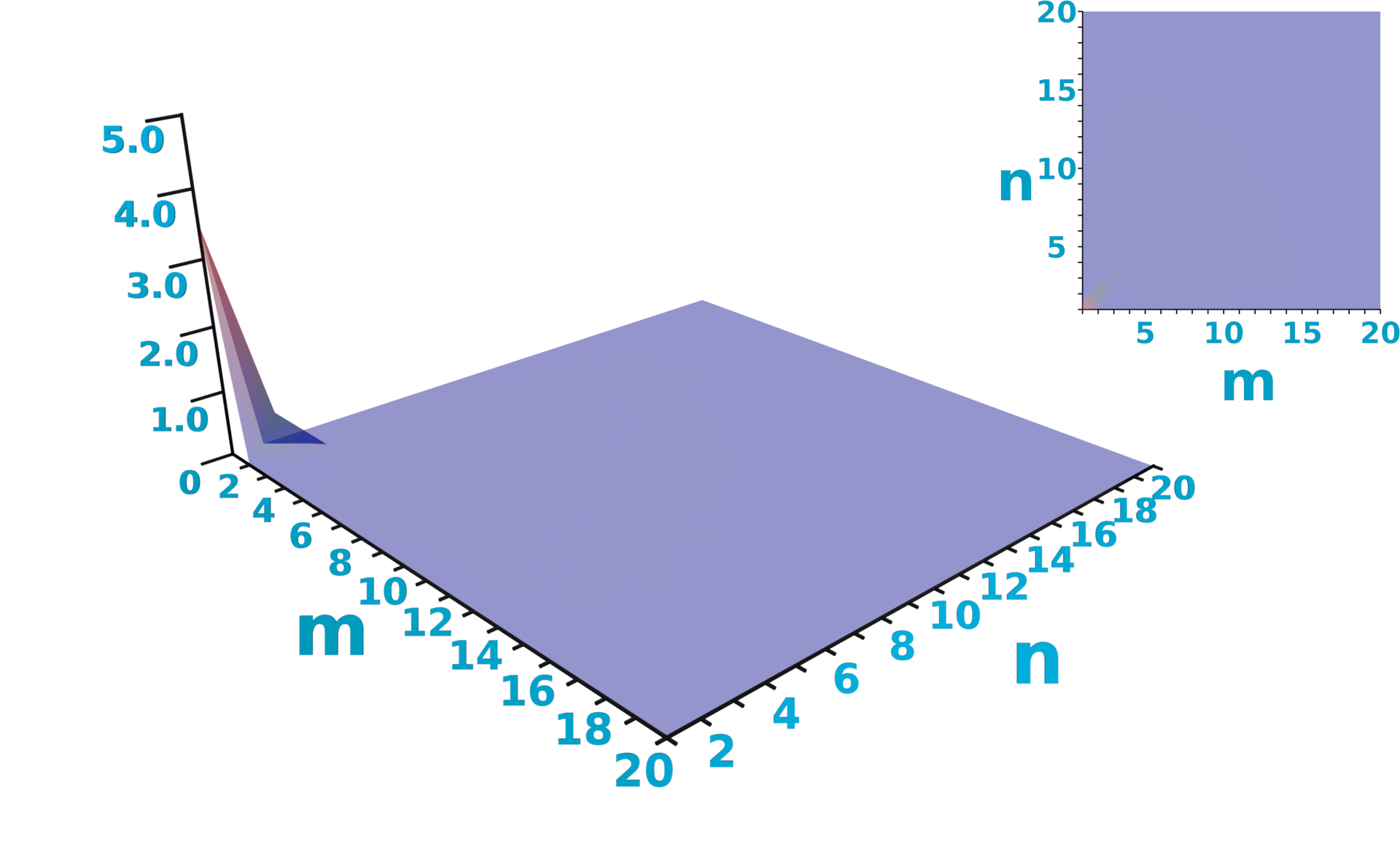}
    }

    \subfloat[][]{
        \label{fig:rholesser3b}
        \includegraphics[width=\linewidth]{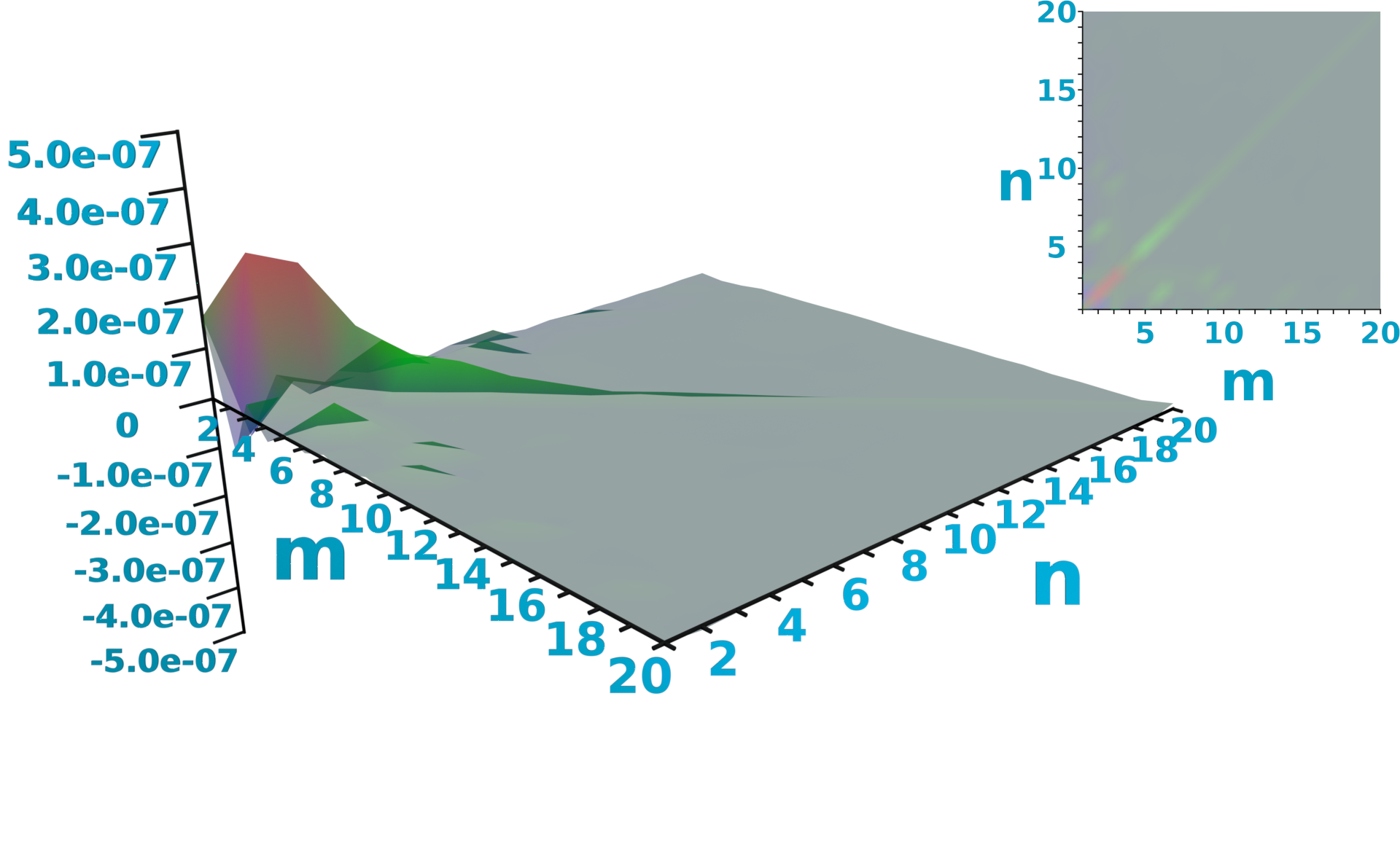}
    }
    \caption[]{Normal and anomalous elliptical magnon densities for a system
    with low anisotropy, $K/J=\num{1e-7}$. All other parameters are equal to
    those in Fig.~\ref{fig:rholesser1}. The horizontal axes represent the
    quantum numbers $m$ and $n$. \subref{fig:rholesser3a}: normal density
    $\big<\psi^\dagger_m(t)\psi_n(t)\big>$.  \subref{fig:rholesser3b}: anomalous
    density $\big<\psi^\dagger_m(t)\psi^\dagger_n(t)\big>$. The minor (light
    green) fluctuations are near the scale of numerical error ($10^{-8}$) and
    may be unphysical.}
    \label{fig:rholesser3}
\end{figure}

\begin{figure}
    \centering
    \subfloat[][]{
        \label{fig:rholesser4a}
        \includegraphics[width=\linewidth]{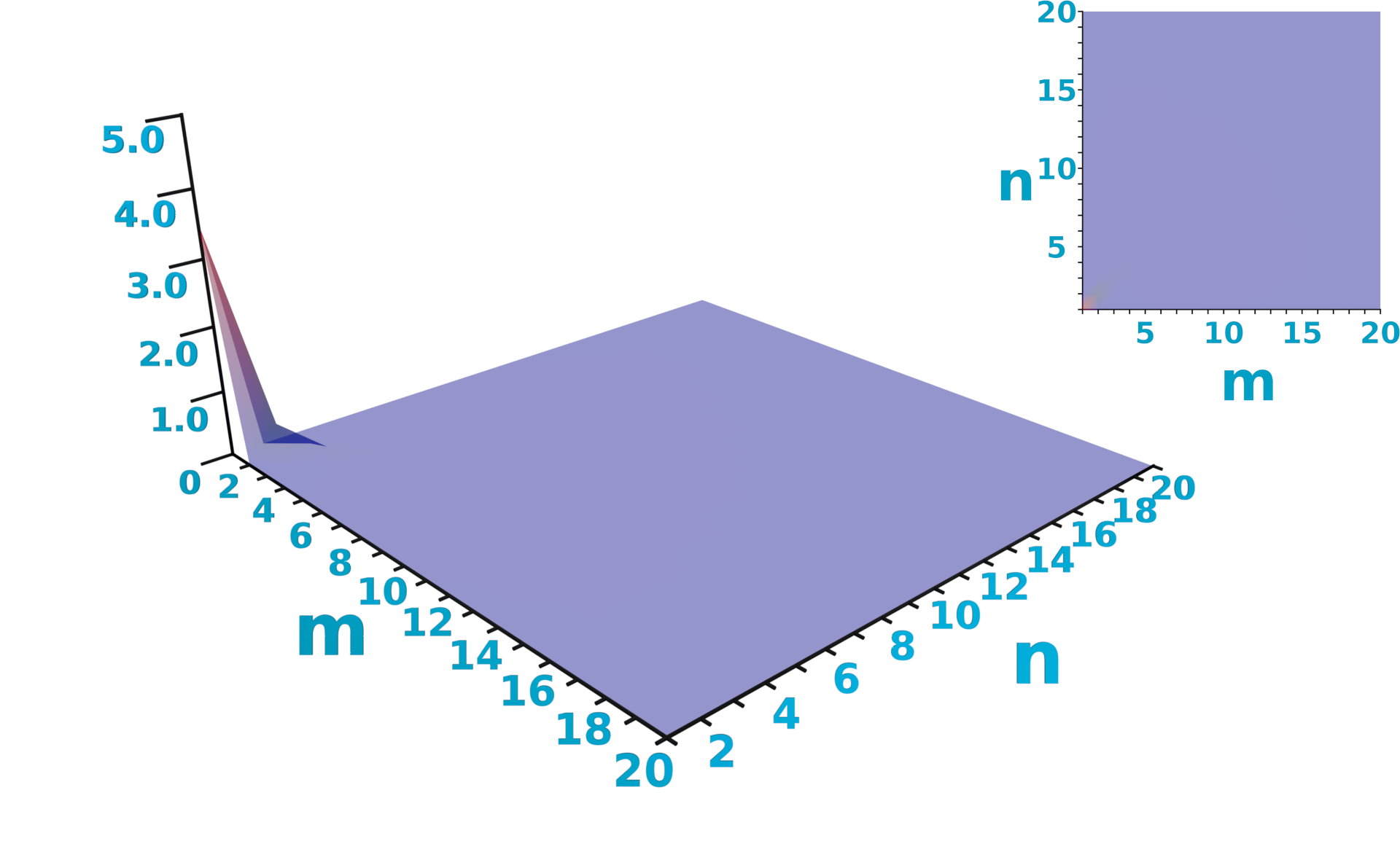}
    }

    \subfloat[][]{
        \label{fig:rholesser4b}
        \includegraphics[width=\linewidth]{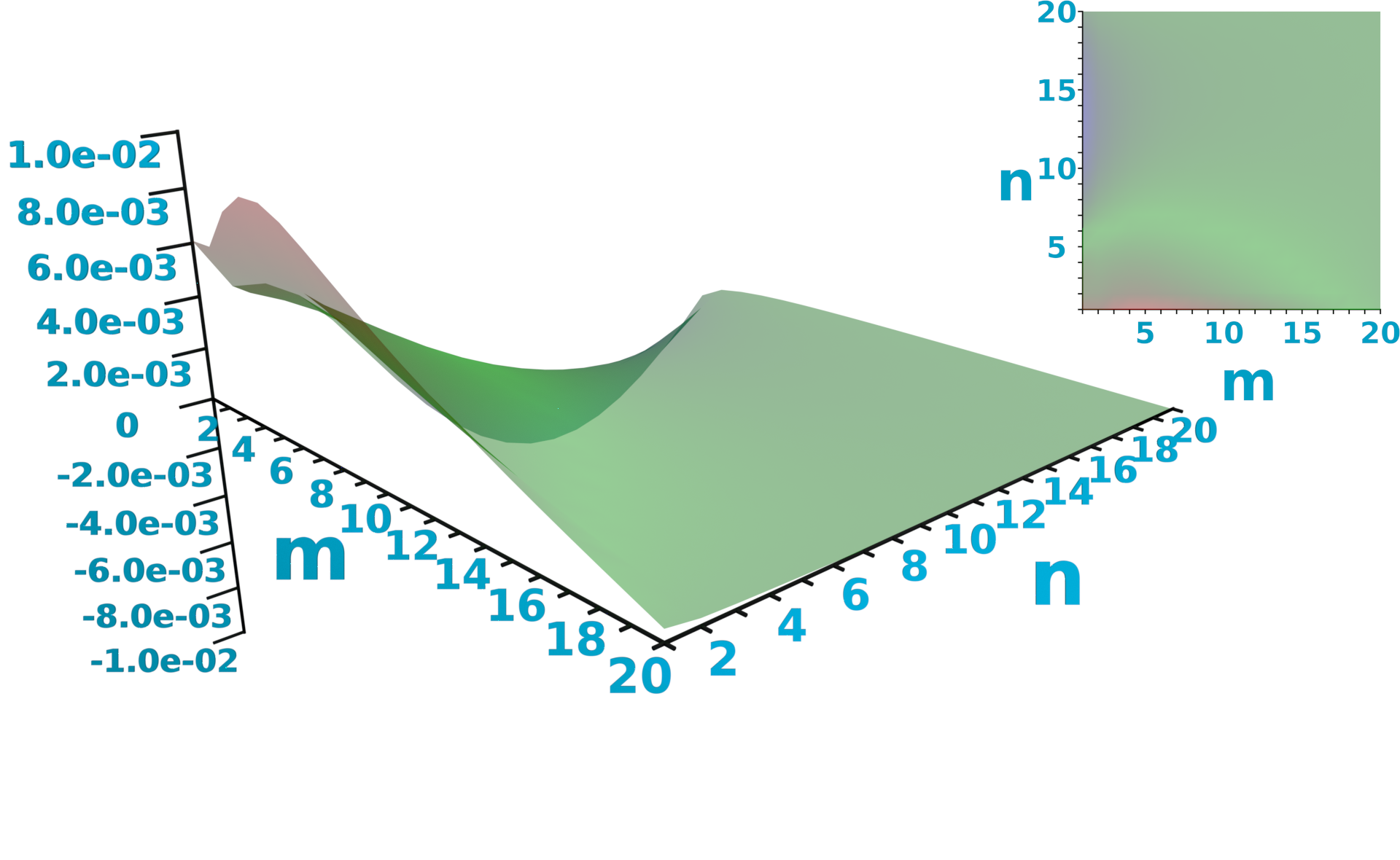}
    }
    \caption[]{Normal and anomalous elliptical magnon densities for a system
    with high anisotropy, $K/J=\num{5e-2}$. All other parameters are equal to
    those in Fig.~\ref{fig:rholesser1}. \subref{fig:rholesser4a}: normal
    density $\big<\psi^\dagger_m(t)\psi_n(t)\big>$.  \subref{fig:rholesser4b}:
    anomalous density $\big<\psi^\dagger_m(t)\psi^\dagger_n(t)\big>$.}
    \label{fig:rholesser4}
\end{figure}

In Figs.~\ref{fig:rholesser3} and~\ref{fig:rholesser4} we plot the equivalent
matrices in the basis of elliptical magnons: the horizontal axes now represent
the quantum number, and the diagonals of the plots are ordered by increasing
energy. In this basis, the ordinary block $\big<\psi^\dagger_m\psi_n\big>$
of the correlation function $\big<\Psi^\dagger_m\Psi_n\big>$ is almost
exactly diagonal at low anisotropy ($K/J=\num{1e-7}$ shown in
Fig.~\ref{fig:rholesser3a}). As we keep the gap $\varepsilon_1$ fixed, our
chosen parameters lead to excitation of the lowest few modes only, regardless
of anisotropy, with the overwhelming majority of quasiparticles being in the
ground state (as indicated by the large spike at $m=n=1$). In
Fig.~\ref{fig:rholesser3b}, it can be seen that the anomalous block
$\big<\psi^\dagger_m\psi^\dagger_n\big>$ nearly vanishes, as expected (the
same is true for $\big<\psi_m\psi_n\big>$).

Figure~\ref{fig:rholesser4a}) shows that the qualitative behavior of the
ordinary correlations $\big<\psi^\dagger_m\psi_n\big>$ does not change
significantly even at the high anisotropy value $K/J=\num{5e-2}$. However, the
anomalous block $\big<\psi^\dagger_m\psi^\dagger_n\big>$, shown in
Fig.~\ref{fig:rholesser4b} now exhibits a small but noticeable deviation from
zero, and becomes asymmetric. This asymmetry ultimately stems from the fact
that $\big<b^\dagger_ib_j\big>\ne\big<b_ib^\dagger_j\big>$. The bosonic
relations are nevertheless preserved because the full matrix
$\big<\Psi^\dagger_m\Psi_n\big>$ is symmetric.

\begin{figure}
    \subfloat[][]{
        \label{fig:squeeze-nobias}
        \includegraphics[width=\linewidth]{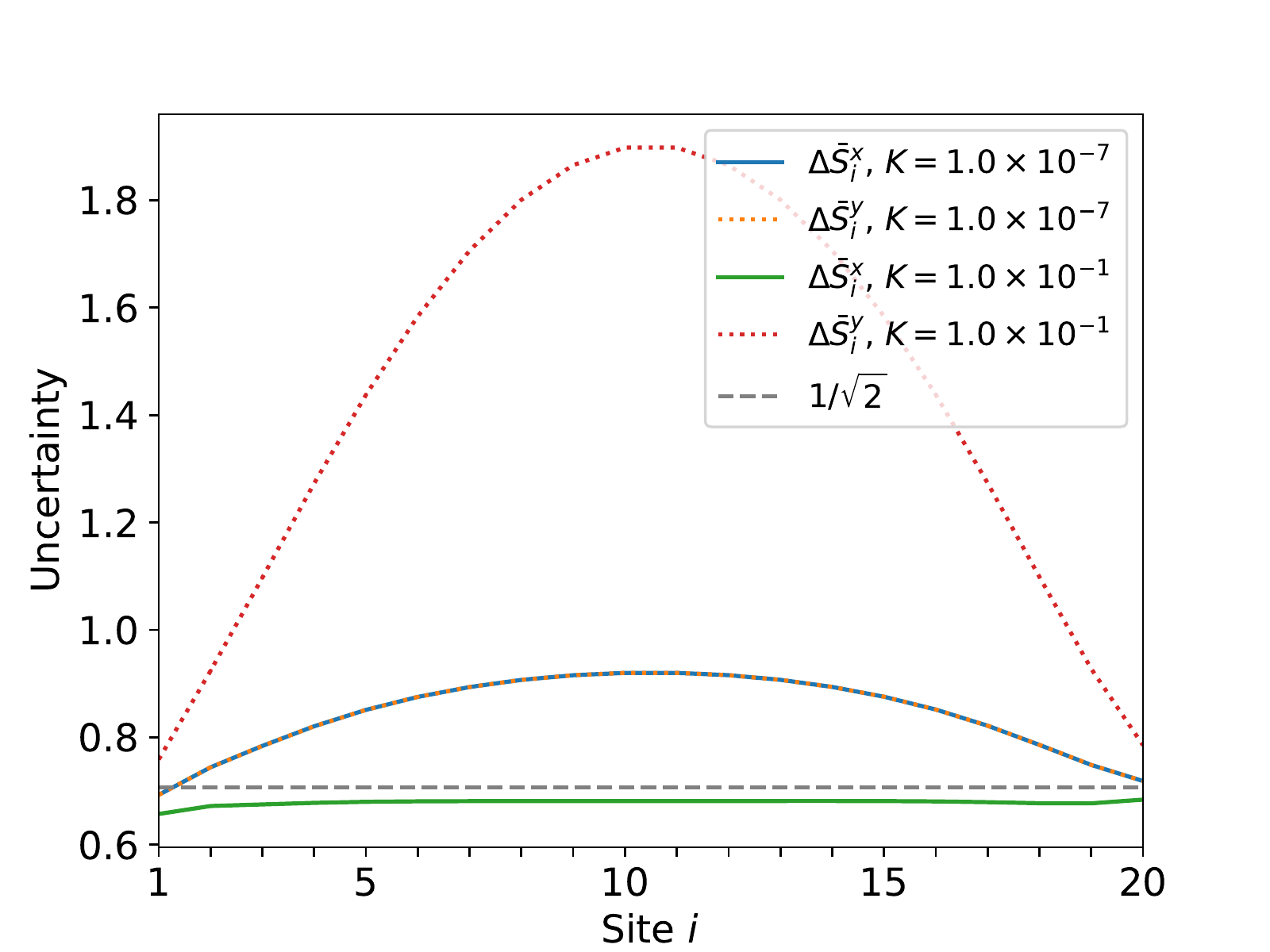}
    }

    \subfloat[][]{
        \label{fig:squeeze-bias}
        \includegraphics[width=\linewidth]{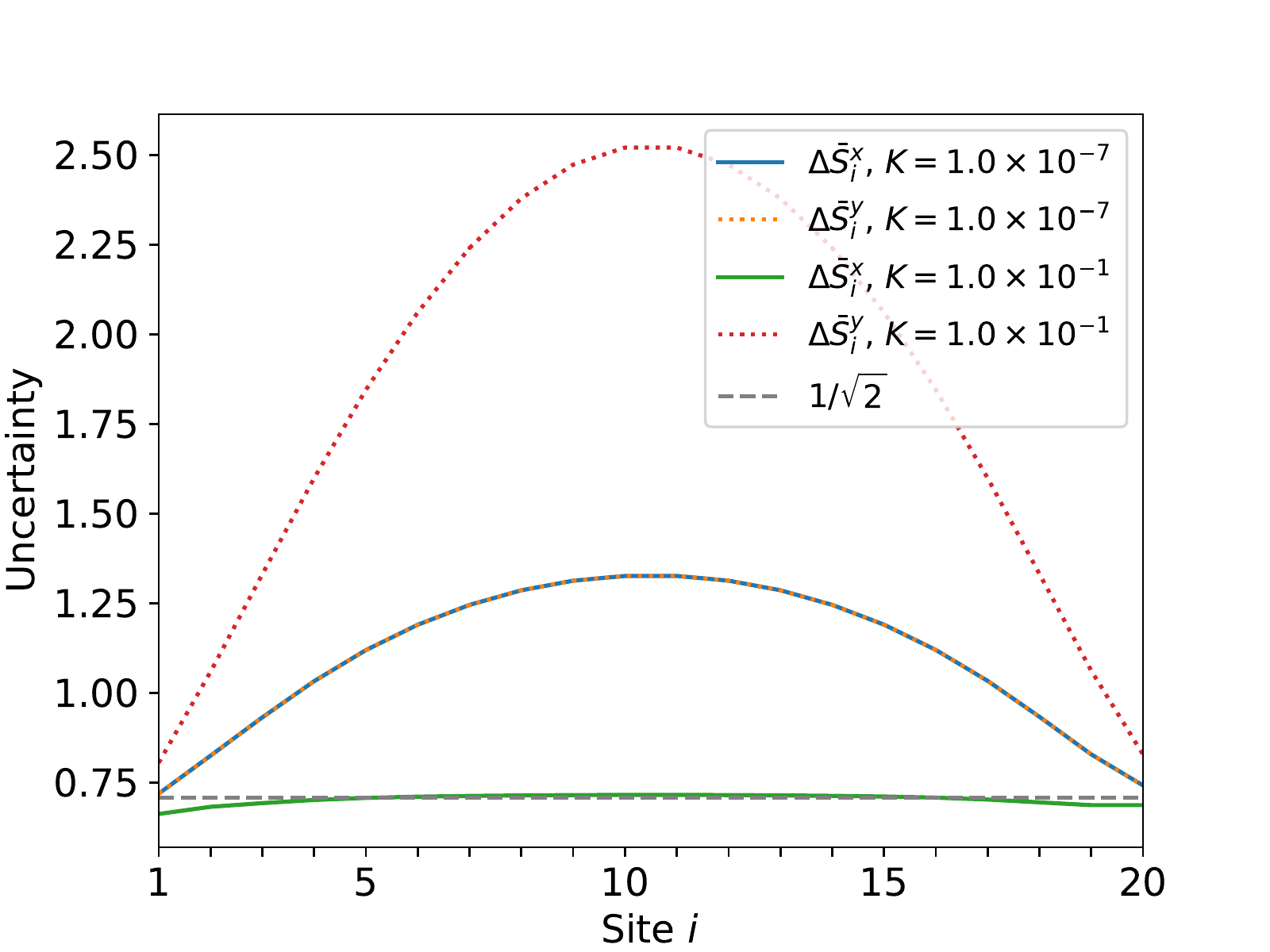}
    }

    \caption[]{Uncertainty in the spin operators $S^x$ (solid lines) and $S^y$
    (dotted lines), for a system of size $N=20$ with only the left lead
    attached ($\eta^\mathrm{L}=0.1$, $\eta^\mathrm{R}=0$), low Gilbert-like
    damping $\alpha=\num{1e-3}$, temperature $\kBT/J=0.1$, gap
    $\varepsilon_1/J=0.025$, and different anisotropies. When the spin
    uncertainty drops below $\frac{1}{\sqrt{2}}$ (dashed line), the state is
    squeezed. At very high anisotropy, the magnons become squeezed throughout
    the system. Note that the $\Delta \bar{S}^x$ and $\Delta \bar{S}^y$ curves
    lie on top of each other at $K/J=\num{1e-7}$, indicating the magnons are
    coherent at low anisotropy.  \subref{fig:squeeze-nobias}: without applied
    bias ($\mu^\mathrm{L}=0$). \subref{fig:squeeze-bias}: with bias at left
    lead ($\mu^\mathrm{L}/J=0.1$).}
    \label{fig:squeeze}
\end{figure}

As explained previously, we may use the density matrix to directly compute
uncertainty of the spin operators, which we expect to become squeezed at high
anisotropy. In Fig.~\ref{fig:squeeze}, we plot the uncertainty amplitudes
$\Delta \bar{S}^x_i$ and $\Delta \bar{S}^y_i$ for $K/J=\num{1e-7}$ and
$K/J=\num{1e-1}$, with weak left-lead coupling $\eta^\mathrm{L}=0.1$ and no
right lead attached. In the case of zero bias ($\mu^\mathrm{L}=0$,
Fig.~\ref{fig:squeeze-nobias}), it can be seen that high anisotropy causes the
magnons to become squeezed throughout the sample. At site 1, where the left
lead is attached, both $\Delta \bar{S}^x$ and $\Delta \bar{S}^y$ are squeezed,
in an apparent violation of the uncertainty principle. However, this may be
explained by the fact that we only consider the lowest-order self-energy
contribution of the lead coupling: this ignores higher-order electronic
contributions to the total wavefunction at the interface, and it stands to
reason---although it remains to be verified---that the uncertainty principle is
not violated if higher-order contributions are taken into account.

Taking only sites $i>1$ into account, we find that squeezing commences at site
20 (the `far side' of the chain, where no lead is attached) for
$K/J\approx\num{3e-2}$. Squeezing increases with increasing anisotropy, with
the effect being strongest at the center of the sample, where the overall spin
density is the highest. By applying a spin bias at the attached lead
($\mu^\mathrm{L}/J=0.1$ shown in Fig.~\ref{fig:squeeze-bias}), squeezing is
diminished throughout the sample, and the overall uncertainty markedly
increases. A local bias may thus be used to effect a global change in the
uncertainty.

\section{Conclusions and outlook}
\label{sec:conc}
We have developed and numerically implemented a NEGF formalism to describe the
transport of elliptically polarized magnons in finite-sized ferromagnetic
insulators terminated by metallic leads. The presence of anisotropy in a
ferromagnetic insulator can give rise to a novel parasitic local spin
resistance, and additionally acts to suppress the spin conductance measured
between the metallic leads. However, our model predicts that these effects are
mild in ferromagnets with weak anisotropy, and become significant only when
the ferromagnet exhibits strong anisotropy.

We have shown that the NEGF formalism allows theoretical access to the
anomalous correlation functions $\big<b_ib_j\big>$ and
$\big<b^\dagger_ib^\dagger_j\big>$, which may obtain a large amplitude in
the presence of anisotropy, and provide a measure for the degree of
nonconservation of spin and ellipticity of magnons. Likewise, the correlation
functions $\big<\psi_m\psi_n\big>$ and
$\big<\psi^\dagger_m\psi^\dagger_n\big>$ in the basis of eigenstates of the
free anisotropic ferromagnetic insulator obtain a nonzero value in the
presence of coupling to metallic leads which inject a well-defined amount of
spin, provided the anisotropy is large and the lead coupling is sufficiently
strong. Moreover, strong anisotropy produces squeezing of $\Delta \bar{S}^x$,
which may be observable in the form of reduced shot noise
\cite{2016PhRvL.116n6601K} and find applications in quantum information
science.

Although we have focussed on ferromagnets, where anisotropy tends to be
significantly weaker than the exchange interaction, it stands to reason that
much stronger observable effects may be realized in antiferromagnets, where
similar anomalous Hamiltonian terms are introduced by coupling between
sublattices, but are now governed by the exchange interaction itself
\cite{2020ApPhL.117i0501K}.

While we have provided some examples of effects produced by the introduction
of anisotropy, our model is simplistic, and omits several features one would
expect to find in a realistic system. A possible extension, for example, would
be the introduction of disorder, which can take the form of spatial
fluctuations in both $\Delta$ and $K$. Moreover, our model considers only weak
interactions between magnons and the leads and lattice (i.e. lowest-order
self-energy terms), while higher-order contributions may be relevant to
physical systems.  We have likewise neglected magnon-magnon interactions,
while several real systems are known or believed to violate this assumption
\cite{1997PhRvB..5515048G,2018PhRvL.120u7202C,2020NatSR..1012548X}.

Finally, the parameter space of our model (with or without extensions) is
quite large, and therefore remains mostly unexplored. Hence, it is plausible
that more observable effects of the spin-conservation breaking anisotropies
can be found, for example through the spin Seebeck effect.

\section{Acknowledgements}
R.A.D. is member of the D-ITP consortium, a program of the Dutch Organisation
for Scientific Research (NWO) that is funded by the Dutch Ministry of
Education, Culture and Science (OCW). This project has received funding from
the European Research Council (ERC) under the European Union’s Horizon 2020
research and innovation programme (grant agreement No. 725509).

A.R. acknowledges financial support by the Deutsche Forschungsgemeinschaft
(DFG) through Project No. KO/1442/10-1.

B.Z.R. acknowledges support by Iran Science Elites Federation (ISEF).

\appendix
\section{Derivation of the steady-state spin current}
\label{app:spincur}
We define the total spin current as the negative time-derivative of the total
HP magnon number density (as each magnon carries spin 1), i.e.\begin{align}
    j_\mathrm{s}^\mathrm{tot}&=-\partial_t\Tr\left<b^\dagger(t) b(t)\right>
            \nonumber \\
        &=-\frac{1}{2}\partial_t\left[\Tr\left\{\left<b^\dagger(t) b(t)\right>+
            \left<b(t)b^\dagger(t)\right>\right\}-N\right] \nonumber \\
        &=-\frac{1}{2}\partial_t\Tr\left<\phi^\dagger(t)\phi(t)\right>
        =-\real\Tr\left<\phi^\dagger(t)\partial_t\phi(t)\right>.
        \label{eq:spincur}
\end{align}
Note that the trace on the first two lines is over the spatial indices alone
and therefore has $N$ terms, whereas on the last line, it is over the full
matrix and therefore has $2N$ terms. To evaluate Eq.~(\ref{eq:spincur}), we
introduce a stochastic field\begin{align} \label{eq:stoch}
    \xi(\omega)&=-g(\omega)\phi(\omega)
\end{align}obeying\begin{align}
    \label{eq:stochexp1}
    \left<\xi(\omega)\xi^\dagger(\omega')\right>
        &=\uppi\upi\delta(\omega-\omega')\Sigma^\mathrm{K}(\omega).
\end{align}

By construction, $\xi$ is the Hubbard-Stratonovich field that decouples the
quantum-quantum term of the Schwinger-Keldysh action for the continuum-limit
field theory. A more detailed derivation is given e.g. by
\citet{kamenev2002keldysh}.

Inserting Eq.~(\ref{eq:dyson}) into Eq.~(\ref{eq:stoch}) and taking the
Fourier transform, we find the evolution equation\begin{align}
    -\upi\partial_t\phi(t)&=\sigma_3h\phi(t)
            +\sigma_3\int\dif t'\Sigma(t-t')\phi(t')-\sigma_3\xi(t),
    \intertext{which we may plug into Eq.~(\ref{eq:spincur}) to obtain}
    j_\mathrm{s}^\mathrm{tot}&=-\imag\Tr\Bigg\{
            \left<\phi^\dagger(t)\sigma_3h\phi(t)\right> \nonumber \\
            &\hspace{6em}+\left<\phi^\dagger(t)\sigma_3\int\dif t'\Sigma(t-t')
                \phi(t')\right> \nonumber \\
            &\hspace{6em}-\left<\phi^\dagger(t)\sigma_3\xi(t)\right>\Bigg\}.
            \label{eq:jssteady}
\end{align}
Here the first term is the Hamiltonian evolution of the system, and the second
and third terms represent driving by external factors: the second term
concerns the interaction with the lead electrons and with the field(s)
responsible for Gilbert-like damping, and the third term contains the effect
of quantum noise.

In the steady state, the total spin current vanishes by definition, and thus
the first term necessarily cancels against the driving terms. In an experiment
where the system is held out of equilibrium by external driving, the net
external source/sink current are then given by the sum of the last two terms
of Eq.~(\ref{eq:jssteady}). However, these terms, as given, sum over all of
the spin currents within the system, including unobservable contributions that
occur deep within the bulk and never exit the ferromagnet, whereas the
actually observable spin currents are those which flow out of the leads. This
quantity is obtained when one replaces $\Sigma$ with $\Sigma^\mathrm{L/R}$ and
$\xi$ with $\xi^\mathrm{L/R}$ in Eq.~(\ref{eq:jssteady}). Here we define
$\xi^\mathrm{L/R}$ to be the stochastic field obeying\begin{align}
    \label{eq:stochexp2}
    \left<\xi^\mathrm{L/R}(\omega)\xi^\dagger(\omega')\right>
            &=2\uppi\upi\delta(\omega-\omega')\Sigma^\mathrm{L/R}(\omega)
                F^\mathrm{L/R}(\omega),
\end{align}
where $2\Sigma^\mathrm{L/R}(\omega)F^\mathrm{L/R}(\omega)$ is the left/right
lead term of the Keldysh self energy.

Thus, focussing now on the spin current flowing out of the left lead (in the
following derivation, one may obtain equivalent expressions for the right lead
by swapping L and R), we find\begin{align}
    j^\mathrm{L,tot}_\mathrm{s}&=-\imag\Tr\Bigg\{
            \left<\phi^\dagger(t)\sigma_3\int\dif t'\Sigma^\mathrm{L}(t-t')
                \phi(t')\right> \nonumber \\
            &\hspace{6em}-\left<\phi^\dagger(t)\sigma_3\xi^\mathrm{L}(t)
                \right>\Bigg\}.
    \intertext{Next, by Fourier transforming and using
    Eq.~(\ref{eq:stoch}) to write $\phi$ in terms of $\xi$, we obtain}
    j^\mathrm{L,tot}_\mathrm{s}&=-\imag\Tr\int\frac{\dif\omega}{2\uppi}
            \frac{\dif\omega'}{2\uppi}\upe^{\upi t(\omega'-\omega)}
                \nonumber \\
        &\hspace{6em}\times\Bigg\{\left<\xi^\dagger(\omega)g^\dagger(\omega)
            \sigma_3\Sigma^\mathrm{L}(\omega')g(\omega')\xi(\omega')\right>
            \nonumber \\
        &\hspace{6em}+\left<\xi^\dagger(\omega)g^\dagger(\omega)
            \sigma_3\xi^\mathrm{L}(\omega')\right>\Bigg\}.
    \intertext{Reordering terms using the properties of the trace and making
    use of Eqs.~(\ref{eq:stochexp1}),~(\ref{eq:stochexp2})
    and~(\ref{eq:sigmaK}), this gives}
    j^\mathrm{L,tot}_\mathrm{s}&=-\real\Tr\int\frac{\dif\omega}{2\uppi}\,
            g^\dagger(\omega)\sigma_3\Sigma^\mathrm{L}(\omega)g(\omega)
            \nonumber \\
        &\hspace{6em}\times\Bigg\{\Sigma^\mathrm{L}(\omega)
            F^\mathrm{L}(\omega)+\Sigma^\mathrm{R}(\omega)F^\mathrm{R}(\omega)
            \nonumber \\
        &\hspace{6em}+\Sigma^\mathrm{B}(\omega)\mathcal{T}^{-1}
                F^\mathrm{B}(\omega)\mathcal{T}
            +g^{-1}(\omega)F^\mathrm{L}(\omega)\Bigg\}.
    \intertext{Inserting the Dyson equation for $g^{-1}(\omega)$, we find}
    j^\mathrm{L,tot}_\mathrm{s}&=-\real\Tr\int\frac{\dif\omega}{2\uppi}\Bigg\{
            \iota^\mathrm{R\to L}(\omega)+\iota^\mathrm{B\to L}(\omega)
            \nonumber \\
        &\hspace{6em}+g^\dagger(\omega)\sigma_3\Sigma^\mathrm{L}(\omega)
            g(\omega)\left[\omega\sigma_3-h\right]F^\mathrm{L}(\omega)\Bigg\},
\end{align}
where $\iota^\mathrm{R\to L}(\omega)$ and $\iota^\mathrm{B\to L}$ are given by
Eqs.~(\ref{eq:iotaR}) and~(\ref{eq:iotaB}), respectively. By using that
$\real\Tr M=\frac{1}{2}\Tr(M+M^\dagger)$ for an arbitrary square matrix $M$,
the term involving $\omega\sigma_3$ can easily be shown to vanish. In a
similar vein, we find\begin{align}
    -\real\Tr&\left\{g^\dagger(\omega)\sigma_3\Sigma^\mathrm{L}(\omega)
            g(\omega)hF^\mathrm{L}(\omega)\right\} \nonumber \\
        &\quad=\frac{1}{2}\Tr\left\{g^\dagger(\omega)\sigma_3
            \Sigma^\mathrm{L}(\omega)g(\omega)
            \left[F^\mathrm{L}(\omega),h\right]\right\}. \label{eq:commut}
\end{align}
In the absence of a spin accumulation, $F^\mathrm{L}(\omega)$ becomes a scalar
function multiplying the identity matrix, causing the commutator to vanish.
Therefore, we may add the term\begin{align}
    0&=\Tr g^\dagger(\omega)\sigma_3\Sigma^\mathrm{L}(\omega)g(\omega)h\left.
            F^\mathrm{L}(\omega)\right|_{\mu^\mathrm{L}=0},
\end{align}
thereby recovering Eq.~\ref{eq:iotaL}). Finally, in the limit $K\to0$, the
Hamiltonian $h$ becomes block-diagonal, so that the commutator in
Eq.~(\ref{eq:commut}) causes $\iota^\mathrm{L}(\omega)$ to vanish in absence
of anisotropy.

\FloatBarrier

\bibliography{refs}

\end{document}